\documentclass[showkeys,showpacs,superscriptaddress,twocolumn,aps,pre,10pt,nofootinbib]{iopart}

\usepackage{amssymb}
\usepackage{graphicx}
\usepackage[caption=false]{subfig}
\usepackage{mathtools}
\usepackage{comment}
\usepackage{cite}
\usepackage{units}
\usepackage[colorlinks=true, allcolors=blue]{hyperref}

\usepackage{cuted}

\begin{document}


\title{A reduced kinetic method for investigating non-local ion heat transport in ideal multi-species plasmas}

\author{N T Mitchell$^1$, D A Chapman$^2$, C J McDevitt$^3$, M P Read$^2$, G Kagan$^1$}

\address{$^1$ The Blackett Laboratory, Imperial College, London SW7 2AZ, UK}

\address{$^2$ First Light Fusion Ltd., Unit 9/10 Oxford Pioneer Park, Mead Road, Yarnton, Kidlington OX5 1QU, UK}

\address{$^3$ Nuclear Engineering Program, Department of Materials Science and Engineering, University of Florida, Gainesville, Florida 32611, USA}

\ead{\mailto{n.mitchell22@imperial.ac.uk}}






\begin{abstract}
    A reduced kinetic method (RKM) with a first-principle collision operator is introduced in a 1D2V planar geometry and implemented in a computationally inexpensive code to investigate non-local ion heat transport in multi-species plasmas. 
    The RKM successfully reproduces local results for multi-species ion systems and the important features expected to arise due to non-local effects on the heat flux are captured. In addition to this, novel features associated with multi-species, as opposed to single species, case are found. Effects of non-locality on the heat flux are investigated in mass and charge symmetric and asymmetric ion mixtures with temperature, pressure, and concentration gradients. In particular, the enthalpy flux associated with diffusion is found to be insensitive to sharp pressure and concentration gradients, increasing its significance in comparison to the conductive heat flux driven by temperature gradients in non-local scenarios. The RKM code can be used for investigating other kinetic and non-local effects in a broader plasma physics context. Due to its relatively low computational cost it can also serve as a practical non-local ion heat flux closure in hydrodynamic simulations or as a training tool for machine learning surrogates.
\end{abstract}

\noindent{\it Keywords\/}: non-local transport, reduced kinetic method, multi-species



\maketitle
\ioptwocol

\allowdisplaybreaks


\section{Introduction}\label{section_intro}

The collisional transport of energy is a significant issue for the design and optimization of a wide range of schemes for realizing nuclear fusion in both magnetized and inertial systems \cite{J_D_Callen_1997}. In the most commonly encountered context of ICF, where a fuel-containing pellet is irradiated directly by lasers \cite{10.1063/1.4934714} or indirectly by a laser-produced thermal X-ray bath \cite{Meezan_2017}, the influence of heat flow is prominent in the transport of energy absorbed from the driver to the ablation front \cite{e_heaT_conduction_icf} and the hot spot. Here, the dominant carriers of the absorbed energy are delocalized (or `free') electrons, which transfer their energy through collisions with ions and other free electrons in the irradiated material. If the temperature gradient is sufficiently shallow, this energy is transported deeper into the unheated ablator through conduction, i.e., the collisional `local' diffusion of heat; the heat flux then obeys Fourier's law \cite{osti_4344773}. However, with high-intensity irradiation the mean free path (MFP) of the high-energy particles that contribute most to the heat flow routinely exceed the scale length of the gradient, resulting in transport with quasi-ballistic features \cite{icf_kinetics_review}. While `non-local' behaviour may be accurately captured using full kinetic simulations \cite{THOMAS20121051}, these methods are impractical for integrated modeling of the implosion process. Alternatively, and more pragmatically, numerous simplified models have been constructed which may readily be implemented as sub-cycled operators in radiation-hydrodynamics codes \cite{10.1063/1.5011818,Holec_2018,Chen_2023,article_brodrick_testing_vfp_snb,10.1063/5.0018733}. At later times, when the fusion fuel is in a state of pre-ignition, heat flow also acts as a major factor in the power balance that determines the pathway and probability of success of fuel burn and the onset of ignition. 

The foregoing arguments are usually framed from the perspective of the electrons, as it is commonly assumed that the ionic heat flux can be neglected in comparison to the electronic contribution. The assumption relies on  estimating the ratio of the ion to electron heat conductivities by the square root of the electron-ion mass ratio. However, as recently discussed by Chapman et al. \cite{Chapman_PhysPlasmas_2021}, for fully ionized fusion fuel one finds parity of the conductivities under modest temperature separations (around $T_{i}/T_{e}\sim 4$ for deuterium), which can be vastly exceeded in burning fusion plasmas. Even more significant separations may be realized during the ignition and burn phase as the hallmark of this process is ion temperature runaway until hydrodynamic decompression occurs and the fuel explodes and rapidly cools. Non-local effects are also expected for ion heat flux since $\alpha$-particle deposition will distort the hot tail of the background distribution function (DF), resulting in an over-abundance of energy-transporting ions relative to a Maxwell-Boltzmann distribution \cite{Atzeni_PhysPlasmas_2007}. 

Hydrodynamic quantities such as density and temperature are populated predominantly by thermal particles in the bulk of the velocity distribution, whereas the heat flux is sourced primarily by suprathermal particles in the tail of the distribution. In a plasma, faster particles have longer MFPs, therefore non-local effects are most apparent in the tail of the DF where particle MFPs may be comparable to length scales on which hydrodynamic quantities vary. This leads to a regime where the bulk is dominated by collisions and near-Maxwellian, leaving hydrodynamic quantities undisturbed, but where the tail of the DF may experience non-local kinetic effects, perturbing the heat flux. A set of fluid equations describing evolution of hydrodynamic quantities in such a regime may then remain a valid description of the plasma, but with a non-local prescription for the heat flux.

In the local limit, ion transport coefficients for a simple plasma of electrons and a single ion species are provided by Braginskii \cite{Braginskii1965ReviewsOP}. Comprehensive formalisms and practical prescriptions for calculations of local ionic transport processes in an unmagnetized plasma with multiple ion species are given by Chapman and Cowling \cite{chapman_cowling}, Ferziger and Kaper \cite{Ferziger_book}, Zhdanov \cite{zhdanov_book}, with compact formulae provided by Kagan and Baalrud \cite{kagan2016transport}, and Kagan and Tang \cite{10.1063/1.4745869,Kagan_2014}. 

The local result for the heat flux is generally a linear combination of gradients of hydrodynamic quantities such as temperature, but extending to other quantities such as concentration and partial pressure in multi-species cases. For a given density and temperature, there is a maximum possible heat flux corresponding to all particles streaming freely at the thermal speed. For a sufficiently sharp gradient, the local result will exceed the free-streaming limit, thereby become unphysical. Non-local effects then have a tendency to inhibit heat flux to remain physical, which is supported by simulations and Fokker-Planck (FP) codes \cite{Epperlein91}. This observation motivates a flux-limiter approach \cite{10.1063/1.1694445,10.1063/1.4921130}, where local transport coefficients are limited in non-local cases by a fitted function compared to full FP methods. Although computationally cheap, flux limiters are physically inaccurate and fail to capture other features of non-local transport such as preheating in cold regions. More advanced models, such as the Schurtz-Nicola{\"i}-Busquet (SNB) approach \cite{SNB_model}, provide a more physically accurate description of non-local transport for electrons but not ions. For ion transport, this leaves flux limiters as the usual practical option implemented in hydrodynamics codes. Chapman et al. \cite{Chapman_PhysPlasmas_2021} indicates that flux limiters may be insufficient for modelling non-local ion heat transport accurately and therefore may be a major contributor to discrepancies in modelling ICF systems of interest, in particular reproducing neutron yields from experiments on uniaxially driven targets \cite{Derentowicz_JourTechPhys_1977, Derentowicz_JourTechPhys_1977b}.

In this work, we present a reduced kinetic method (RKM) based on approaches by McDevitt et al. \cite{McDevitt_1, McDevitt_2, McDevitt_3} to describe non-local behaviour of the ion DFs, and therefore ionic heat flux, in a multi-species plasma. The RKM solves only for the tail of the DF where non-local effects are expected to be most pronounced. Further noting that fewer particles reside on the tail compared to the bulk, a linearized Coulomb collision operator is valid even for order-unity deviations from a Maxwellian distribution in the tail. 
Transport quantities such as the heat flux can then be computed from the DFs provided by the RKM solver. This leaves a novel first-principles model for non-local ion transport, which as of now has not been widely studied even for a simple, single ion species plasma. Importantly, this study is conducted in the multi-species context where ionic heat flux includes additional features not present in the single ion species or electron cases. In particular, an additional contribution to the heat flux exists due to relative diffusion of different ion species which is generally of comparable magnitude to the standard conductive component. Multi-species scenarios also allow gradients of partial pressures and concentrations act as drives for transport fluxes along with the standard drive due to the temperature gradient. The impact of non-locality on both of these multi-species effects is for the first time investigated. We consider two components of the heat flux, the reduced heat flux and the enthalpy flux, and find that non-local deviations of the latter can be insensitive to sharp pressure and concentration gradients, therefore increasingly important in comparison to the reduced heat flux in non-local scenarios. Many hydrodynamics codes treat ion mixtures as a single effective species, thereby ignore the enthalpy component of the heat flux which may be particularly significant with sharp concentration gradients and therefore influence fluid simulations. 

The remainder of this paper is organized as follows; in \ref{section_rkm}, the mathematical model as derived from the VFP equation and numerical implementation is presented. In \ref{section_heat_flux}, the additional complexity of multi-species heat flux in comparison to the single-species case is described. In \ref{section_results}, results from the RKM are presented for a single species and weakly and strongly asymmetric ion mixtures with gradients in temperature, pressure, and concentration, as well as consistency checks, convergence studies, and comparisons to SNB and flux-limited models.



\section{Reduced kinetic method}\label{section_rkm}

Ideal, weakly coupled plasmas may be described by the VFP equation, which is the Boltzmann equation for a plasma with a Fokker-Planck collision operator as a consequence of long-range Coulomb collisions with multiple species. The VFP equation for a species $\alpha$ in laboratory-frame phase-space coordinates $(t,\boldsymbol{x},\boldsymbol{v})$ for particle position $\boldsymbol{x}$ and particle velocity $\boldsymbol{v}$ is

\begin{equation}
    \label{VFP_full}
    \partial_t f_\alpha  + \boldsymbol{v}\cdot\boldsymbol{\nabla}f_\alpha + \boldsymbol{F}_\alpha\cdot\boldsymbol{\nabla}_{\boldsymbol{v}}f_\alpha  = \sum_\beta C_{\alpha\beta} \{ f_\alpha , f_\beta\}
    \,,
\end{equation}
where $f_\alpha$ is the DF of species $\alpha$ and $\boldsymbol{F}_\alpha$ is the force per unit mass acting on species $\alpha$. The FP collision operator $ C_{\alpha\beta} \{ f_\alpha , f_\beta \}$ represents collisions of species $\beta$ acting on species $\alpha$, with the sum over all species $\beta$ including $\alpha$. Ion-electron collisions are almost always negligible compared to ion-ion collisions and are therefore neglected, although they may be readily included.

It is convenient to decompose each species' DF into a Maxwellian component and its deviation from Maxwellian as $f_\alpha = f^M_\alpha + \delta f_\alpha$. A Maxwellian distribution of a species is given by $f^M_\alpha = n_\alpha / ( \sqrt{\pi} v_{\text{th},\alpha} )^{3} \exp( - (\vert \boldsymbol{v} - \boldsymbol{u} \vert / v_{\text{th},\alpha})^2 )  $, where $n_\alpha$ is the species' number density, $T_\alpha$ is the species' temperature in units of energy, and $v_{\text{th},\alpha} \equiv \sqrt{2T_\alpha/m_\alpha}$ is the species' thermal velocity with particle mass $m_\alpha$. The centre-of-momentum velocity, or hydrodynamic velocity, is defined as $\boldsymbol{u} = \sum_\alpha c_\alpha \boldsymbol{u}_\alpha$, where $\boldsymbol{u}_\alpha = \frac{1}{n_\alpha}\int d^3\boldsymbol{v}\ \boldsymbol{v}f_\alpha$ is the the net flow velocity of a species $\alpha$. The difference between these different velocities and the hydrodynamic velocity is typically much smaller than their thermal speeds $\vert \boldsymbol{u}_\alpha - \boldsymbol{u} \vert \ll v_{\text{th},\alpha}$ for an ion mixture near local thermal equilibrium, therefore the Maxwellian drift velocity for all species may be chosen to be $\boldsymbol{u}$ instead of $\boldsymbol{u}_\alpha$ for convenience. Although different species may have their own temperatures in the mathematical framework, ion species with comparable masses and different temperatures is indicative of a system far from local thermal equilibrium, therefore a common ion temperature $T_\alpha = T$ is taken. Using bilinearity of the FP operator, and noting that Maxwellians satisfy $C_{\alpha\beta}\{f^M_\alpha,f^M_\beta\} = 0 $ for a common temperature and net flow velocity, the collision operator can be decomposed as

\begin{equation}
    \label{coll_op}
    \begin{aligned}
        C_{\alpha\beta} \{ f_\alpha , f_\beta \} 
         = &\,
         C_{\alpha\beta} \{ \delta f_\alpha , f^M_\beta \} 
         + 
         C_{\alpha\beta} \{ f^M_{\alpha} , \delta f_\beta \} 
         \\& + 
         C_{\alpha\beta} \{ \delta f_{\alpha} , \delta f_\beta \} 
    \end{aligned}   
\end{equation} 

The first term on the RHS of (\ref{coll_op}) $C_{\alpha\beta}^T\{ \delta f_\alpha\} = C_{\alpha\beta} \{ \delta f_\alpha , f^M_\beta \} $ is the `test-particle' operator which describes collisions of species $\alpha$ against a Maxwellian background of species $\beta$. The second term $C_{\alpha\beta}^F\{ \delta f_\beta\} = C_{\alpha\beta} \{ f^M_{\alpha} , \delta f_\beta \}$ is the `field-particle' operator which describes the effect of species $\beta$ on the Maxwellian part of species $\alpha$. It can be argued that the third, non-linear term $C_{\alpha\beta} \{ \delta f_{\alpha} , \delta f_\beta \}$ is subleading to the test-particle term $\vert C_{\alpha\beta} \{ \delta f_{\alpha} , \delta f_\beta \} \vert \ll \vert C^T_{\alpha\beta} \{ \delta f_{\alpha} \} \vert $ even for order-unity deviations from Maxwellian in the tail of the distribution, since $\delta f_\beta (\boldsymbol{v}') $ is smaller than $f^M_\beta (\boldsymbol{v}')$ globally in $\boldsymbol{v}'$ space, and the second argument of $C_{\alpha\beta}$ is integrated over $\boldsymbol{v}'$ \cite{Landau_1936}. Hence, the full kinetic equation is reduced to a linear problem with the collision operator

\begin{equation}
    \label{coll_op_linearised}
    \begin{aligned}
        C_{\alpha\beta} \{ f_\alpha , f_\beta \} 
         \approx &\,
         C_{\alpha\beta}^{T}\{\delta f_\alpha\}
         + 
         C_{\alpha\beta}^{F} \{\delta f_\beta \}
         \,,
    \end{aligned}   
\end{equation} 
which is numerically tractable while still capturing relevant kinetic physics.


\begin{figure}
    \centering
    \includegraphics[width=0.49\textwidth]{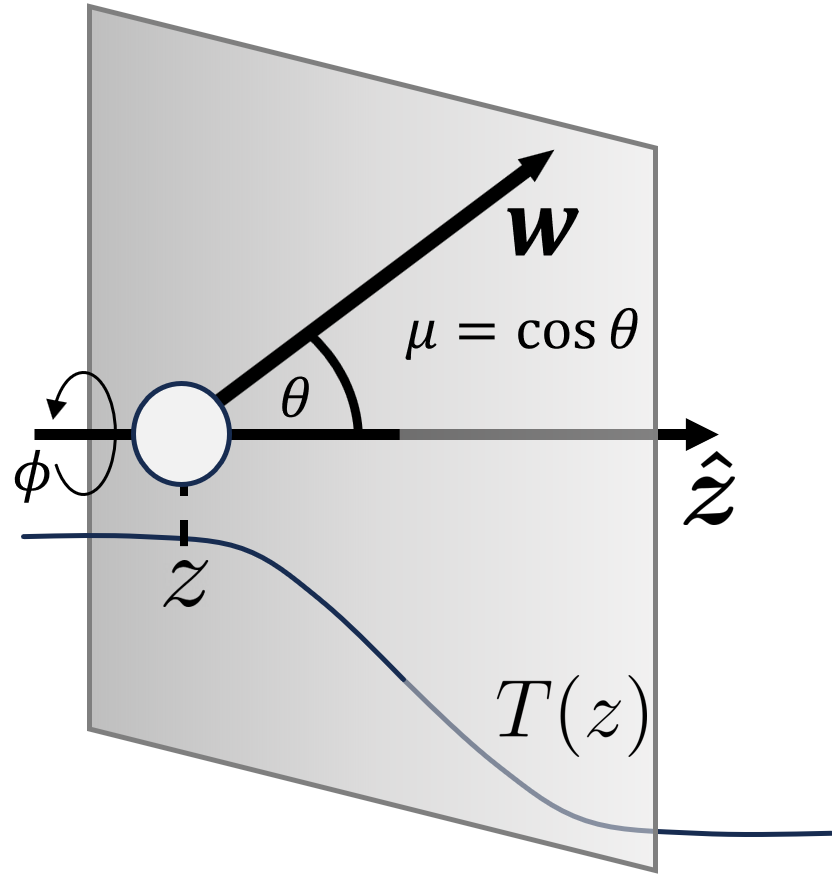}
    \caption{
        Visualization of 1D2V geometry at a boundary. There is an azimuthal symmetry in the velocity coordinate $\phi$, and translational symmetry parallel to the boundary.
    }
    \label{1d2V_sketch}
\end{figure}

It is often the case that the length scale of a boundary in a fluid system is much shorter than its spatial extent in the other two directions parallel to the boundary, therefore the geometry near the boundary may be considered to be near one-dimensional planar (1D). This requires that any scalar quantities of the problem (such as temperature and number densities) only vary in the direction perpendicular to the boundary. In a one-dimensional spatial planar geometry, there is an azimuthal symmetry around the single spatial direction, therefore the velocity space becomes two dimensional (2V), as depicted in figure \ref{1d2V_sketch}.

Transport quantities such as heat flux are defined in the centre-of-momentum frame. \ref{appendix_kinetic_equation_for_deviation} gives the kinetic equation for the deviation of the DF from Maxwellian $\delta f_\alpha = f_\alpha - f^M_\alpha$ in the centre-of-momentum frame in a 1D2V planar geometry as

\begin{equation}
    \label{kinetic_equation_for_deviation_section2}
    \begin{aligned}
        \partial_t \delta f_\alpha &=  \sum_\beta \bigg( C^T_{\alpha\beta} \{ \delta f_\alpha \} + C^F_{\alpha\beta} \{ \delta f_\beta \} \bigg) - \mu w \partial_z \delta f_\alpha \\&- \bigg( \frac{1}{\rho}\partial_z p + F_\alpha - \sum_\beta c_\beta F_\beta \bigg) 
         \bigg( \mu \partial_w  + \frac{1-\mu^2}{w}\partial_\mu  \bigg)\delta f_\alpha 
        \\&-\mu w f^M_\alpha  \bigg(\frac{n}{n_\alpha} d_\alpha + \bigg( \frac{m_\alpha }{2T}w^2 -\frac{5}{2}\bigg)\partial_z \ln T  \bigg)
        \,,
    \end{aligned}
\end{equation}
where $z$ is the single planar spatial coordinate, $\boldsymbol{w} = \boldsymbol{v} - \boldsymbol{u}$ is the velocity in the centre-of-momentum frame (peculiar velocity), $w \equiv \vert \boldsymbol{w} \vert$ is particle speed in the centre-of-momentum frame, and $\mu \equiv \boldsymbol{w}\cdot \boldsymbol{\hat{z}} / w = \cos \vartheta$ is the cosine of the velocity pitch angle $\vartheta$ in the centre-of-momentum frame. The diffusive driving force $d_\alpha$ is defined in Ferziger and Kaper \cite{Ferziger_book} as $d_\alpha = (\partial_z p_\alpha - c_\alpha \partial_z p)/p - c_\alpha \rho / p(F_\alpha - \sum_\beta c_\beta F_\beta)$. The concentration $c_\alpha = \rho_\alpha / \rho$ is the ratio of the species' mass density $\rho_\alpha = n_\alpha m_\alpha$ to the total mass density of the mixture $\rho = \sum_\beta \rho_\beta$. The partial pressure of species $\alpha$ is given as $p_\alpha = n_\alpha T$, and $p = \sum_\alpha p_\alpha$ is the total pressure of the ion mixture. The hydrodynamic velocity has been neglected by assuming it is much smaller than all thermal speeds $\vert \boldsymbol{u} \vert \ll v_{\text{th},\alpha}$. (\ref{kinetic_equation_for_deviation_section2}) will then be relaxed to $\partial_t \delta f_\alpha = 0$ since we solve for transport quantities being functions only of the instantaneous hydrodynamic profiles.

\subsection{Test-particle operator}

The test-particle collision operator against a Maxwellian background \cite{helander_and_sigmar} of species $\beta$ may be written in the form
\begin{equation}
    \label{C_alpha_beta}
    C^T_{\alpha\beta}\{ f_\alpha \} = \nu_D^{\alpha\beta} \hat{\mathcal{L}}_\mu f_\alpha + \frac{1}{w^2} \partial_w \bigg( \tilde{\nu}_s^{\alpha\beta} w^3 f_\alpha + \frac{1}{2}\nu_\parallel^{\alpha\beta} w^4 \partial_w f_\alpha \bigg)
    \,,
\end{equation}
where $\hat{\mathcal{L}}_\mu f = \frac{1}{2}\partial_\mu[(1-\mu^2)\partial_\mu f]$ is the pitch-angle scattering operator with azimuthal symmetry, and the three collision frequencies are defined via

\begin{equation}\label{collision_frequencies}
    \begin{aligned}
         \nu^{\alpha\beta}_D &= \hat{\nu}_{\alpha\beta}\frac{\text{erf}(\xi_\beta) - G(\xi_\beta)}{\xi_\alpha^3},
         \\ \tilde{\nu}_s^{\alpha\beta} &= \frac{m_\alpha}{m_\alpha + m_\beta} \nu^{\alpha\beta}_s = 2\hat{\nu}_{\alpha\beta}\frac{G(\xi_\beta)}{\xi_\alpha},
        \\ \nu_\parallel^{\alpha\beta} &= 2\hat{\nu}_{\alpha\beta}\frac{G(\xi_\beta)}{\xi_\alpha^3}
        \,,
    \end{aligned}
\end{equation}
where $\xi_\alpha = w / v_{\text{th},\alpha}$, $\text{erf}(x)$ is the error function \cite{abramowitz+stegun}, $G(x) = [\text{erf}(x) - \frac{2}{\sqrt{\pi}}xe^{-x^2} ]/2x^2 $ is the Chandrasekhar function, and the basic collision frequency is

\begin{equation}
    \hat{\nu}_{\alpha\beta} = \frac{n_\beta Z_\alpha^2 Z_\beta^2 e^4 \ln\Lambda_{\alpha\beta}}{8\pi \sqrt{2m_\alpha} \varepsilon_0^2  T^{3/2}_\alpha}
    \,,
\end{equation}
where we take $\ln\Lambda_{\alpha\beta} = \ln\Lambda$ as the conventional Coulomb logarithm for simplicity. The latter may take many different forms as determined from theoretical arguments \cite{Spitzer1962} or fits to high-fidelity plasma microphysics simulations, however throughout the remainder of this work the Coulomb logarithm will be treated as a tunable parameter of the RKM framework. With the foregoing considerations the test-particle collision operator \eqref{C_alpha_beta} is then linear in species $\beta$, such that $C^T_{\alpha} = \sum_\beta C^T_{\alpha\beta}$ can be written in the more compact form

\begin{equation}
    C^T_\alpha \{ \delta f_\alpha \} 
    = 
    \nu_D^{\alpha} \hat{\mathcal{L}}_\mu \delta f_\alpha 
    + \frac{1}{w^2} \partial_w \bigg( \tilde{\nu}_s^{\alpha} w^3 \delta f_\alpha 
    + \frac{1}{2}\nu_\parallel^{\alpha} w^4 \partial_w \delta f_\alpha \bigg)
    \,,
\end{equation}
where $\nu^\alpha = \sum_\beta \nu^{\alpha\beta}$ for each of the three collision frequencies $\nu_D^\alpha$, $\Tilde{\nu}^\alpha_s$, and $\nu_\parallel^\alpha$. This collision operator appears similar to the single species form (i.e., if the collision operator \eqref{C_alpha_beta} had only a single species present), but where the collision frequencies may have a more complicated $v$ dependence in asymmetric mass and charge cases due to the error and Chandrasekhar functions appearing in (\ref{collision_frequencies}).

\subsection{Field-particle operator}\label{field_particle_operator}

Rosenbluth, MacDonald, and Judd \cite{PhysRev.107.1,10.1063/1.4936799} give the field-particle operator in 1D2V coordinates as 

\begin{equation} \begin{aligned}
    &C^F_{\alpha\beta}\{ \delta f_\beta\} =  4\pi \bigg(\frac{q_\alpha q_\beta}{m_\alpha} \bigg)^2 \ln \Lambda_{\alpha\beta}\frac{m_\alpha f^M_\alpha}{T_\alpha}  \Biggl[ \frac{4\pi T_\alpha}{m_\beta} \delta f_\beta \\& - \mathcal{H}_\beta 
    + \bigg(\frac{m_\alpha}{m_\beta} - 1 \bigg) w \partial_w \mathcal{H}_\beta + \frac{m_\alpha w^2}{2 T_\alpha} \partial_w^2 \mathcal{G}_\beta \Biggl],
\end{aligned}\end{equation} 
where the Rosenbluth potentials $\mathcal{H}_\beta$ and $\mathcal{G}_\beta$ and their appropriate derivatives are given in terms of $\delta f_\beta$ as Legendre series over $\mu$ as described in \ref{appendix_field_particle_operator}.

The test-particle operator contains only velocity derivatives on $\delta f_\alpha$, therefore is local in velocity space and does not couple the deviations of different species. In contrast, the field-particle operator contains integrals over velocity space and acts on $\delta f_\beta$, therefore is non-local in velocity space and couples the deviations of different species. This makes the field-particle operator more difficult to include analytically and numerically than the test-particle operator, yet retaining the field-particle operator is required for agreement with analytic results in the local limit such as Braginskii's heat flux results, as well as for other important features such as momentum conservation. 

Particles in the tail mainly experience collisions with the bulk of the distribution since it is the most populated region in velocity space. Since the tail is located far away in velocity space from the bulk, tail particles do not resolve the non-Maxwellian part of the bulk well and therefore effectively collide with a Maxwellian background. Such collisions are represented by the test-particle operator, therefore it is expected that the test-particle operator dominates over the field-particle operator in the tail of the distribution, which is the region of phase space which we are primarily interested in for non-local behaviour. This feature allows for a tractable numerical implementation of the field-particle operator as discussed in \ref{numerical_implementation}.

\subsection{Expansion in Legendre polynomials}

Similarly to previous work \cite{PhysRevLett.46.243,McDevitt_1}, $\delta f_\alpha$ may be expanded in Legendre polynomials as

\begin{equation}
    \delta f_\alpha(z,w,\mu) 
    = 
    \sum_{l=0}^\infty \delta f_\alpha^{(l)}(z,w) P_l(\mu)
    \,,
\end{equation}
where the Legendre components
\begin{equation}
    \delta f^{(l)}_\alpha(z,w) 
    = \frac{2}{2l+1}\int_{-1}^1 d\mu\ P_l(\mu) \delta f_\alpha(z,w,\mu) 
    \,
\end{equation}
which follows from orthogonality of Legendre polynomials. Legendre polynomials are a natural basis to expand in over $\mu$ since they are the azimuthally symmetric versions of spherical harmonics which provide a complete orthogonal basis over the unit sphere. Integrating (\ref{kinetic_equation_for_deviation_section2}) over $\frac{2}{2m+1}\int_{-1}^1 d\mu\ P_m(\mu)$ and making use of the orthogonality, recursion relations, and eigenvalue property $\hat{\mathcal{L}}_\mu P_m = - \frac{1}{2}m(m+1) P_m$ of Legendre polynomials yields the set of PDEs

\begin{equation}\label{f_l_PDEs}
\begin{aligned}
    \partial_t \delta f^{(l)}_\alpha 
    & = C_\alpha^{T,(l)}\{\delta f_\alpha^{(l)} \} + \sum_\beta C_{\alpha\beta}^{F,(l)}\{\delta f_\beta^{(l)} \}
    \\& - \biggl(w\partial_z + \mathcal{F}_\alpha \partial_w \biggl)  \biggl( \frac{l}{2l-1}  \delta f^{(l-1)}_\alpha + \frac{l+1}{2l+3}  \delta f^{(l+1)}_\alpha \biggl)
    \\& - \frac{1}{w} \mathcal{F}_\alpha \biggl( \frac{l(l-1)}{2l-1}  \delta f^{(l-1)}_\alpha - \frac{(l+2)(l+1)}{2l+3}  \delta f^{(l+1)}_\alpha \biggl)
     \\& -  w f^M_\alpha  \bigg(\frac{n}{n_\alpha} d_\alpha + \bigg( \frac{m_\alpha }{2T}w^2 -\frac{5}{2}\bigg)\partial_z \ln T  \bigg) \delta^{l,1}
    \,,
\end{aligned}
\end{equation}
where $\mathcal{F}_\alpha = \frac{1}{\rho}\partial_z p + F_\alpha - \sum_\beta c_\beta F_\beta$, the form of $C_{\alpha\beta}^{F,(l)}\{\delta f_\beta^{(l)} \}$ is given in \ref{appendix_field_particle_operator}, and $\delta^{l,k}$ is the Kronecker delta.

Truncating the expansion to the first $N_l$ Legendre modes, the set of PDEs may be solved numerically. Fine detail of $\delta f_\alpha$ in the pitch angle $\mu$ is not expected, so the series should converge for a reasonable number of Legendre modes. This is checked with a convergence study over $N_l$, as well as having been demonstrated in previous work by McDevitt et al. \cite{McDevitt_2} for a similar problem by comparing the solutions of the system of PDEs for the Legendre components to the full solution of the 3D equation without expanding in Legendre polynomials over $\mu$. 

\begin{figure}
    \centering    \includegraphics[width=0.49\textwidth]{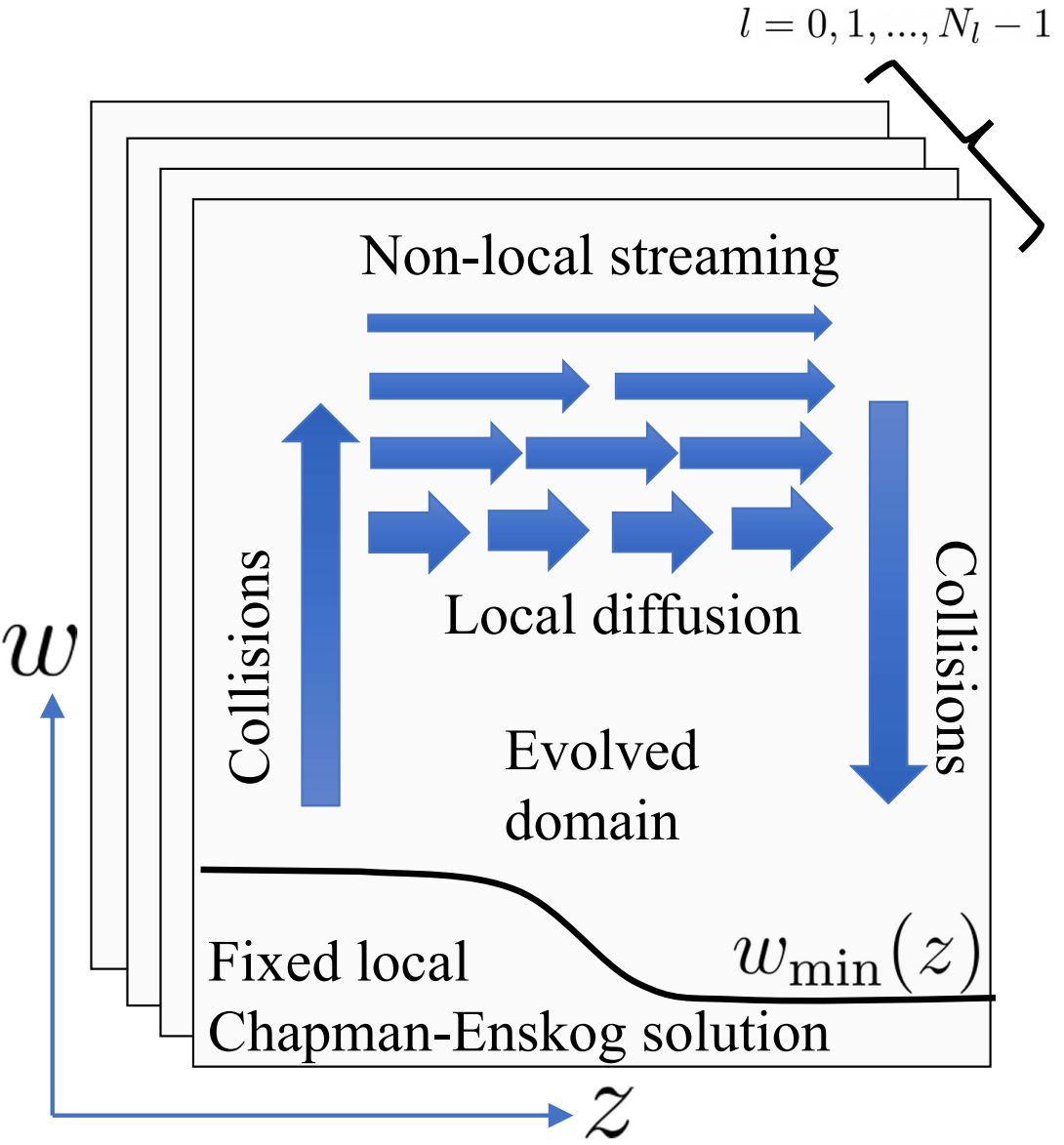}
    \caption{
        Visualization of solver domain for the set of PDEs for $\delta f_\alpha^{(l)}$.
    }
    \label{domain_diagram}
\end{figure}

\subsection{Numerical implementation}\label{numerical_implementation}

The set of PDEs (\ref{f_l_PDEs}) is solved by discretizing $\delta f_\alpha^{(l)}(z,w)$ for all species together to a 4D array $F_{pq}^{s,l}$, where $z = p \Delta z$, $w = q \Delta w$, $l$ is the Legendre index, and $s$ is the species index. Derivatives are estimated via the finite differences

\begin{equation}
    \begin{aligned}
        \partial_{z} F_{p,q}^{s,l} = (F_{p+1/2,q}^{s,l} -F_{p-1/2,q}^{s,l} ) / \Delta z
        \\ 
        \partial_{w} F_{p,q}^{s,l} = (F_{p,q+1/2}^{s,l} -F_{p,q-1/2}^{s,l} ) / \Delta w
        \,,
    \end{aligned}
\end{equation}
where the cell-face values are interpolated via a linear scheme
\begin{equation}
    \begin{aligned}
        F_{p+1/2,q}^{s,l} 
        = &\,
        (1-\delta^z_{p+1/2,q})F_{p+1,q}^{s,l} + \delta^z_{p+1/2,q} F_{p,q}^{s,l}
        \\ 
        F_{p,q+1/2}^{s,l} 
        = &\,
        (1-\delta^w_{p,q+1/2})F_{p,q+1}^{s,l} + \delta^w_{p,q+1/2} F_{p,q}^{s,l}
        \,.
    \end{aligned}
\end{equation}

The coefficients $\delta^z_{p,q}$ and $\delta^w_{p,q}$ (which are functions of grid position) specify the linear interpolation scheme. $\delta^z_{p,q} = 1/2$ is chosen for a spatial central difference and a `Chang-Cooper'-inspired \cite{CHANG_COOPER} scheme $\delta^w_{p,q} = 1/u - 1/(e^u - 1)$, where $u = 2 (\Delta w)^2 q / v_{\text{th},p}^2$ is chosen for velocity derivatives. This choice improves stability and convergence with respect to a simpler finite difference scheme with fixed $\delta^w_{p,q}$. Reshaping the mesh $F_{pq}^{s,l}$ into a vector $\boldsymbol{F}$, the set of PDEs is discretized to $\partial_t \boldsymbol{F} = M (\boldsymbol{F}) + \boldsymbol{P}$, where $M$ is the linear operator representing the discretized right side of (\ref{f_l_PDEs}) acting on $\delta f^{(l)}_\alpha$, and $\boldsymbol{P}$ is the last inhomogeneous term reshaped as a vector. The stationary solution may be found via an explicit or implicit relaxation method. Here, SciPy's backward-differentiation formulas implicit integrator \cite{2020SciPy-NMeth, 10.1145/355626.355636} is used to relax to the steady-state solution.

As discussed in \ref{field_particle_operator}, the field-particle term is negligible in the tail compared to the test-particle operator. This is computationally convenient since the Jacobian of the field-particle operator is non-sparse in the velocity index, whereas the Jacobian of the test-particle operator is. For the solver, the RHS of (\ref{f_l_PDEs}) without the non-sparse integral terms appearing in the field-particle operator may then be used as an approximate Jacobian, with the full field-particle operator entering along with the full timestep function. Although the field-particle operator requires computing $\Psi_{\beta,(1,2,3,4)}^{(l)}(w)$ integrals as defined in \ref{appendix_field_particle_operator} for each $l$, species $\beta$, and spatial position, the integrals are cumulative in velocity, therefore may be computed with reasonable efficiency (see \ref{appendix_field_particle_operator}). Alternatively, the Jacobian of the field-particle term can be included, and these two approaches have been found to be in good agreement.

The leading-order Chapman-Enskog (CE) solution \cite{chapman_cowling,Ferziger_book} is chosen for the the boundary conditions for the DF at spatial edges (where spatial gradients become shallow, therefore transport should become local) and at the low tail cutoff ($w_{\text{min}}= \xi_{\text{min}} v_{\text{th},\alpha}$, with $\xi_{\text{min}} \sim 1$). The spatial boundary conditions are motivated physically by arguing that over a sufficient length, hot particles responsible for non-local effects will have had many collisions and will not penetrate much further away from the boundary, so the local limit should be restored. The local solution is exponentially small for the choice of functions of the bulk profiles, so choosing $f=f^M_\alpha$ as a spatial boundary condition is nearly identical. However, the local perturbative solution does provide a useful boundary condition for the tail cutoff, since $\delta f_\alpha$ would otherwise be zero for $w\leq w_{\text{min}}$, which would remove possibly important parts of the velocity integrand when calculating fluxes. Since the bulk particles have shorter MFPs, we still expect the bulk to match the local solution. At some maximum velocity cutoff above which contributions to the heat flux integrands are expected to be negligible, $\delta f_\alpha=0$ is chosen due to the fast exponential decay. 

\section{Heat flux in a multi-species plasma}\label{section_heat_flux}

Considering the heat flux in a multi-species plasma has further complications in comparison to the single species case. A single species may only experience self-collisions whereas multi-species transport requires also considering collisions with other species. Furthermore, different species in mixtures diffuse in the centre-of-momentum frame which gives rise to an additional source of heat transport of generally comparable magnitude. These additional complications in the multi-species case are also apparent mathematically since there is a set of coupled kinetic equations to solve instead of one in the single species case. Experimental evidence for various types of ICF implosions indicates that multi-species effects can directly affect the yield \cite{PhysRevLett.108.075002,PhysRevLett.112.135001,PhysRevLett.112.185001,PhysRevLett.105.115005}, further motivating the present study.

For a plasma with $N$ ion species, there are $N$ coupled kinetic equations which evolve each species' DF. From this set of kinetic equations, a set of $5N$ fluid equations may be derived for each species' mass density, flow velocity, and temperature. Since the timescales on which different ion species exchange momentum and energy are much shorter than other (often hydrodynamic) timescales of interest, the ions typically have a single common temperature and similar mean flow. There may be cases with highly asymmetric mass mixtures where such an approximation is not valid, but these are not considered in the present work. Such extensions could be considered in the future if the need arises. The separate temperature evolution equations may be combined to form a single evolution equation for the common ion temperature, where the heat flux enters only by the species-summed total heat flux $\boldsymbol{Q} = \sum_\alpha \boldsymbol{Q}_\alpha$. 

The heat flux for a species $\alpha$ is defined in the centre-of-momentum frame as $\boldsymbol{Q}_\alpha = \int d^3\boldsymbol{w}\ \frac{1}{2}m_\alpha w^2 \boldsymbol{w} f_\alpha$. In a simple plasma featuring electrons and a single ion species, there is no heat flux associated with the flow since the only ion species velocity is identical to the hydrodynamic velocity. On the other hand, in a multi-species plasma, each species $\alpha$ has an individual diffusive flow defined as $\boldsymbol{V}_\alpha = \boldsymbol{u}_\alpha - \boldsymbol{u}$ relative to the hydrodynamic velocity. This introduces a component to the heat flux in multi-species cases due to the diffusion of different species in the centre-of-momentum frame transporting energy, which does not occur in the single-species case. The heat flux $\boldsymbol{Q}$ may be decomposed into two parts via $\boldsymbol{Q} = \boldsymbol{q} + \boldsymbol{h}$, the reduced heat flux and enthalpy flux respectively, defined as 

\begin{equation}
    \begin{aligned}
        \boldsymbol{q} 
        =&\, 
        \sum_\alpha \int d^3\boldsymbol{w}\ \biggl( \frac{1}{2} m_\alpha w^2 - \frac{5}{2}T \biggl) \boldsymbol{w} f_\alpha
        \,,
        \\
        \boldsymbol{h} 
        = &\,
        \sum_\alpha \frac{5}{2}T  \int d^3\boldsymbol{w}\   \boldsymbol{w} f_\alpha
        \,.
    \end{aligned}
\end{equation}

Although it is not necessary to decompose $\boldsymbol{Q}=\boldsymbol{q}+\boldsymbol{h}$, it provides more points of comparison for the RKM with the analytic result in the local limit.

The CE expansion \cite{chapman_cowling,Ferziger_book} of multi-species Boltzmann equations yields local results for transport quantities, in particular the heat fluxes in the centre-of-momentum frame
\begin{equation} \begin{aligned}\label{q_ferziger}
    \boldsymbol{q}^\text{local} = - \lambda' \boldsymbol{\nabla}T - nT  \sum_\alpha D^{(T)}_\alpha \boldsymbol{d}_\alpha
    \,,
\end{aligned}\end{equation} 
and
\begin{equation} \begin{aligned}
    \label{h_ferziger}
    \boldsymbol{h} = \frac{5}{2} \sum_\alpha n_\alpha T \boldsymbol{V}_\alpha
    \,,
\end{aligned}\end{equation} 
in which
\begin{equation} \begin{aligned}
    \label{V_ferziger}
    \boldsymbol{V}_\alpha^\text{local} = - D^{(T)}_\alpha \boldsymbol{\nabla}\ln T - \sum_\beta D_{\alpha\beta}\boldsymbol{d}_\beta
    \,, 
\end{aligned}\end{equation} 
where $\lambda'$, $D^{(T)}_\alpha$, and $D_{\alpha\beta}$ are the partial coefficient of thermal conductivity, multi-species thermo-diffusion coefficients, and multi-species (ordinary) diffusion coefficients respectively, with the sign convention following Ferziger and Kaper \cite{Ferziger_book}. These transport coefficients are determined from the collision operator $C_{\alpha\beta}$. For the full Coulomb collision operator with Debye shielding, these may be provided analytically. Here the weakly coupled limit results of \cite{kagan2016transport} are used, which provide numerical routines for the multi-species coefficients.

In the local single ion species case, the only two gradients available to induce heat flux are the temperature and pressure gradients. By assuming force balance, the single-species ion heat flux may be expressed purely in terms of the temperature gradient. In contrast, the multi-species case involves more gradients on which the heat flux may be dependent, such as concentration gradients. In addition, for single species cases the force $\boldsymbol{F}$ does not appear explicitly in the kinetic equation for $\delta f$ due to force balance. For multi-species cases, the force $\boldsymbol{F}_\alpha$ may vary for different species, such as different specific charges $e Z_\alpha / m_\alpha$, giving different electrostatic forces on each species, therefore these forces will appear explicitly in the kinetic equation for $\delta f_\alpha$ and therefore affect the heat flux.

In 1D2V coordinates, the diffusive velocities in the $\boldsymbol{\hat{z}}$ direction are $n_\alpha V_\alpha = \frac{4\pi}{3}\int_0^\infty dw\ w^3 \delta f^{(1)}_\alpha$, and the heat fluxes are

\begin{equation}
    \begin{aligned}
        q 
        = &\, 
        \frac{2\pi}{3}\, \sum_\alpha m_\alpha \int_0^\infty dw\ w^3
        \biggl( w^2 - \frac{5}{2}v_{\text{th}}^2 \biggl)  f_\alpha^{(1)},
    \end{aligned}
\end{equation}
and
\begin{equation}
    \begin{aligned}
        h_\alpha &= \sum_\alpha \frac{5}{2}n_\alpha T V_\alpha
        \,,
    \end{aligned}
\end{equation}
which vanishes for a single species since $u_\alpha = u$.





\section{Results}\label{section_results}

\subsection{Single species case}\label{subsection_single_species}

Consider a single species with a number density profile $n(z)$ and a temperature step of the form

\begin{equation}
    \label{temperature_error_function}
    T(z) 
    = 
    \frac{T_{\text{hot}} + T_{\text{cold}}}{2} + \frac{T_{\text{hot}} - T_{\text{cold}}}{2}
    \text{erf}\left(\dfrac{z}{L}\right)
    \,,
\end{equation}
which transitions from $T_{\text{hot}}$ to $T_{\text{cold}}$ with increasing $z$ over some spatial scale $L$. $n_0 = n_{\text{total}}(z=-\infty)$, $T_0 = T(z=-\infty)$, $m_0 = \min_\alpha (m_\alpha)$, and $v_0 = \sqrt{2T_0 / m_0} $ are introduced as the characteristic number density, temperature, mass, and velocity to which quantities are appropriately non-dimensionalized. In particular, heat fluxes are normalized to a characteristic heat flux $Q_c = n_0 T_0 v_0$, DFs are normalized to $n_0 v_0^{-3}$, and operators appearing in the kinetic equation are normalized to a characteristic collision frequency $\nu_0 = n_0 e^4 \ln \Lambda / 4\pi \epsilon_0^2 m_0^2 v_0^3$. The Knudsen number is introduced as $N_K = \lambda_0 / L$ as a measure of non-locality, where $\lambda_0 = v_0 / \nu_0 = 16\pi \varepsilon_0^2 T_0^2 / n_0 e^4 \ln\Lambda$ is the MFP of a thermal particle at $z=-\infty$. The RKM solver is run for a temperature step $T_{\text{hot}} / T_{\text{cold}} = 10$ and a variety of Knudsen numbers from the local limit to non-local cases. The electric field has been set to zero to focus on the non-local effects due to the temperature profile, although it can be readily included.

\begin{figure}[!htpb]
  \centering
  \subfloat[Isobaric case]{%
    \includegraphics[width=0.49\textwidth]{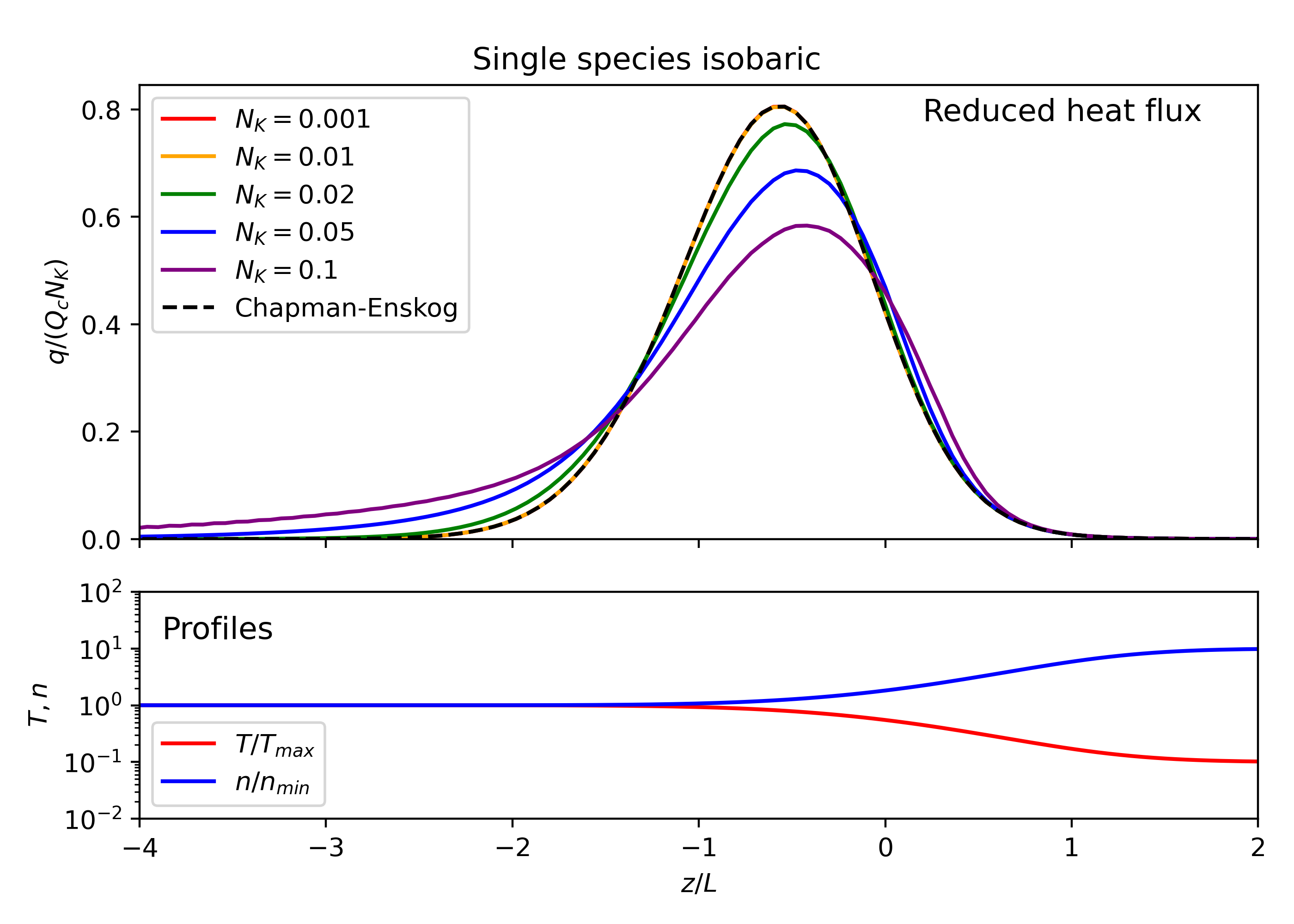}%
  }\\
  \subfloat[Isochoric case]{%
    \includegraphics[width=0.49\textwidth]{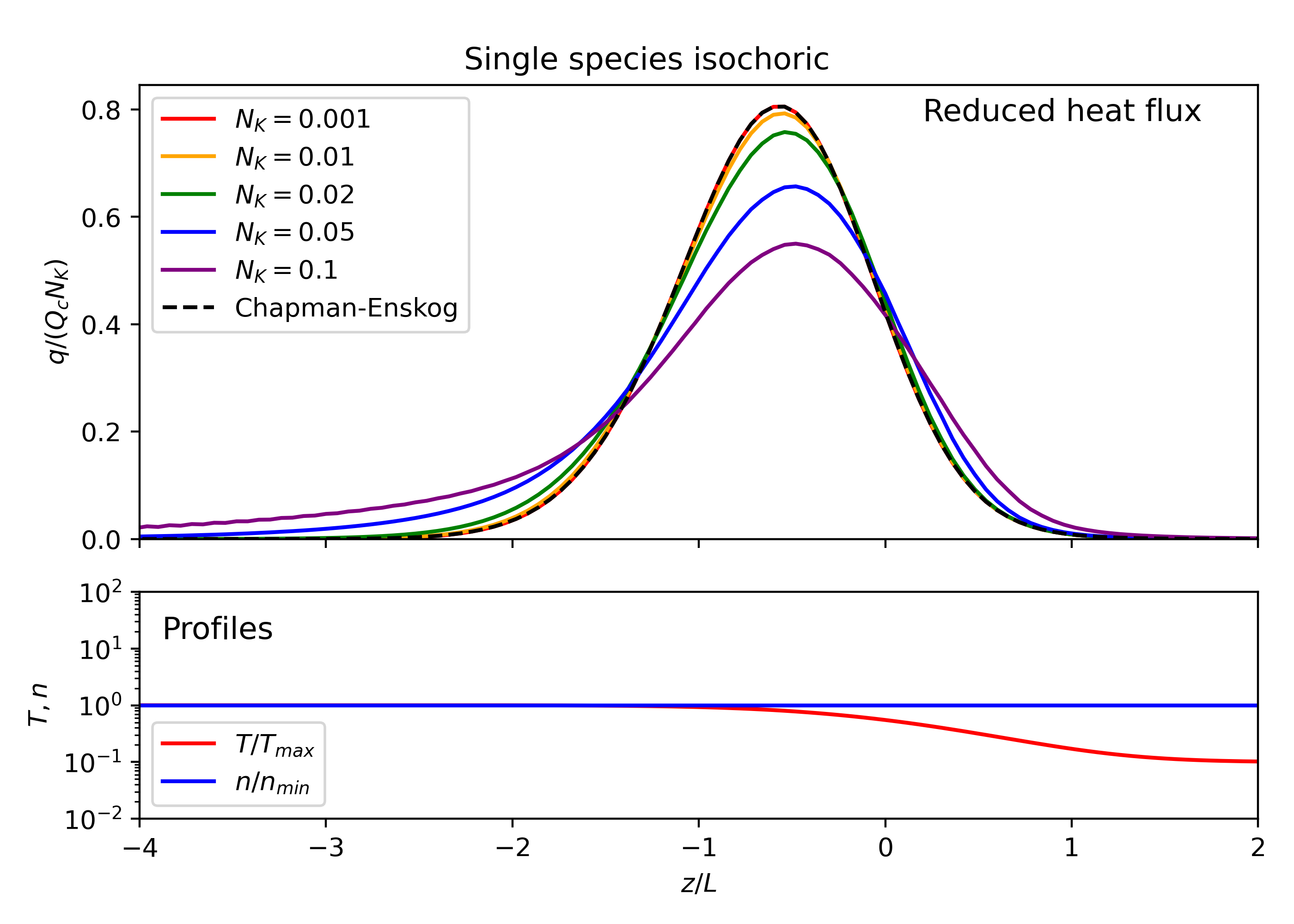}%
  }
  \caption{Reduced heat flux $q$ (rescaled by the Knudsen number, $N_K$) for a single species isobaric (a) and isochoric (b) temperature step for a variety of Knudsen numbers from the RKM solver. The local analytic result for the reduced collision operator and Braginskii's exact result for the full Coulomb collision operator are included. The enthalpy flux has not been included since it is identically zero for a single species, and the temperature and number density profiles are shown below on the same horizontal scale for reference.}
  \label{single_species_fluxes}
\end{figure}

Figure \ref{single_species_fluxes} shows the reduced heat flux for an isobaric ($p=nT=\text{constant}$) and isochoric ($n=\text{constant}$) temperature step. In the local limit the two cases produce the same result and agree well with the local Braginskii result. The latter is also the same in the isochoric and isobaric case. In non-local cases, the isochoric heat flux forms a pedestal in the cold end, whereas the isobaric heat flux does not. This is expected since the higher density in the cold end of the isobaric case gives a higher collisionality, therefore fast particles streaming from the hot end do not penetrate as far.

\begin{figure}[!htpb]
  \centering
  \subfloat[$f^{(0)}$ in hot end]{%
    \includegraphics[width=0.49\textwidth]{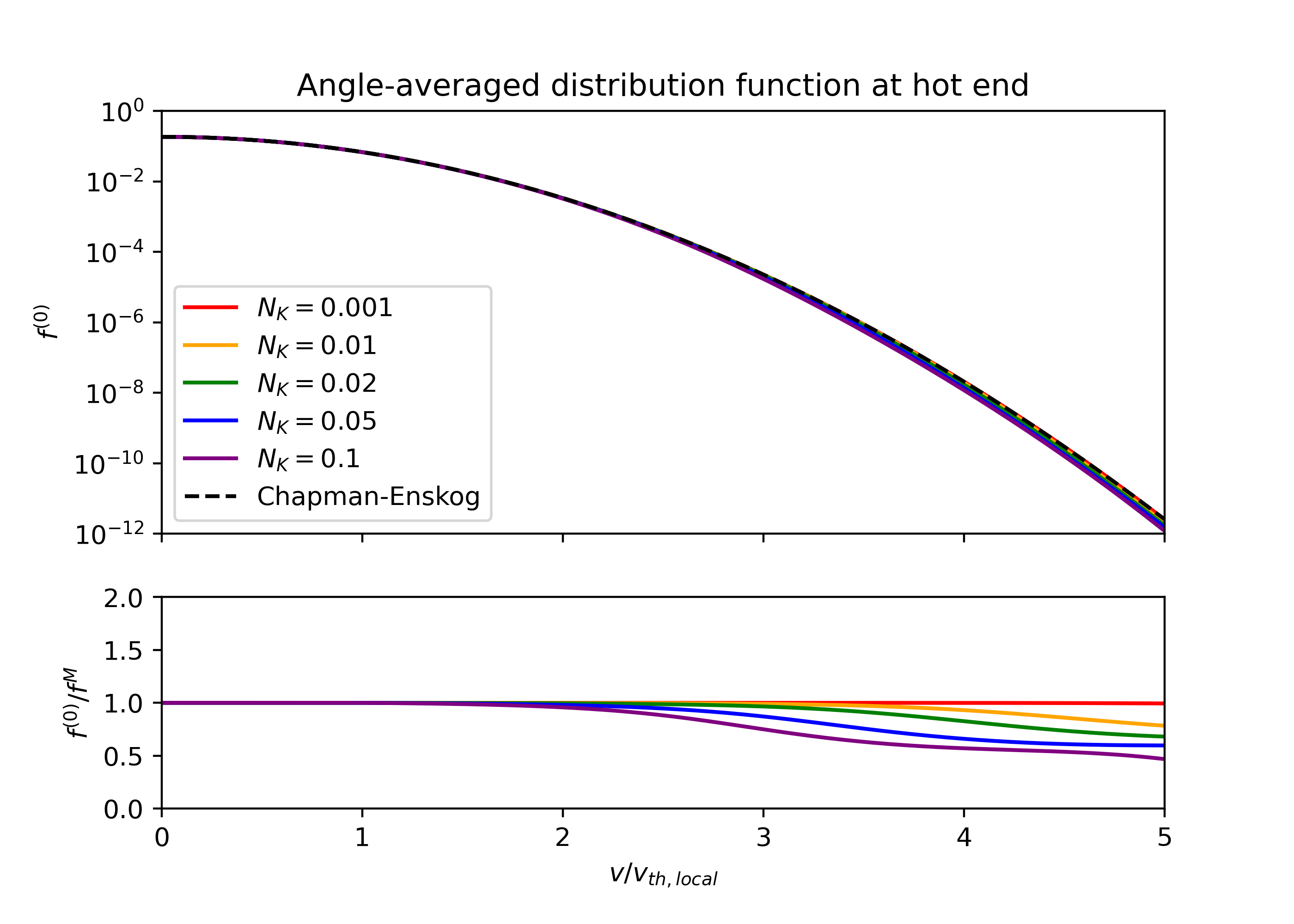}%
  }\\
  \subfloat[$f^{(0)}$ in cold end]{%
    \includegraphics[width=0.49\textwidth]{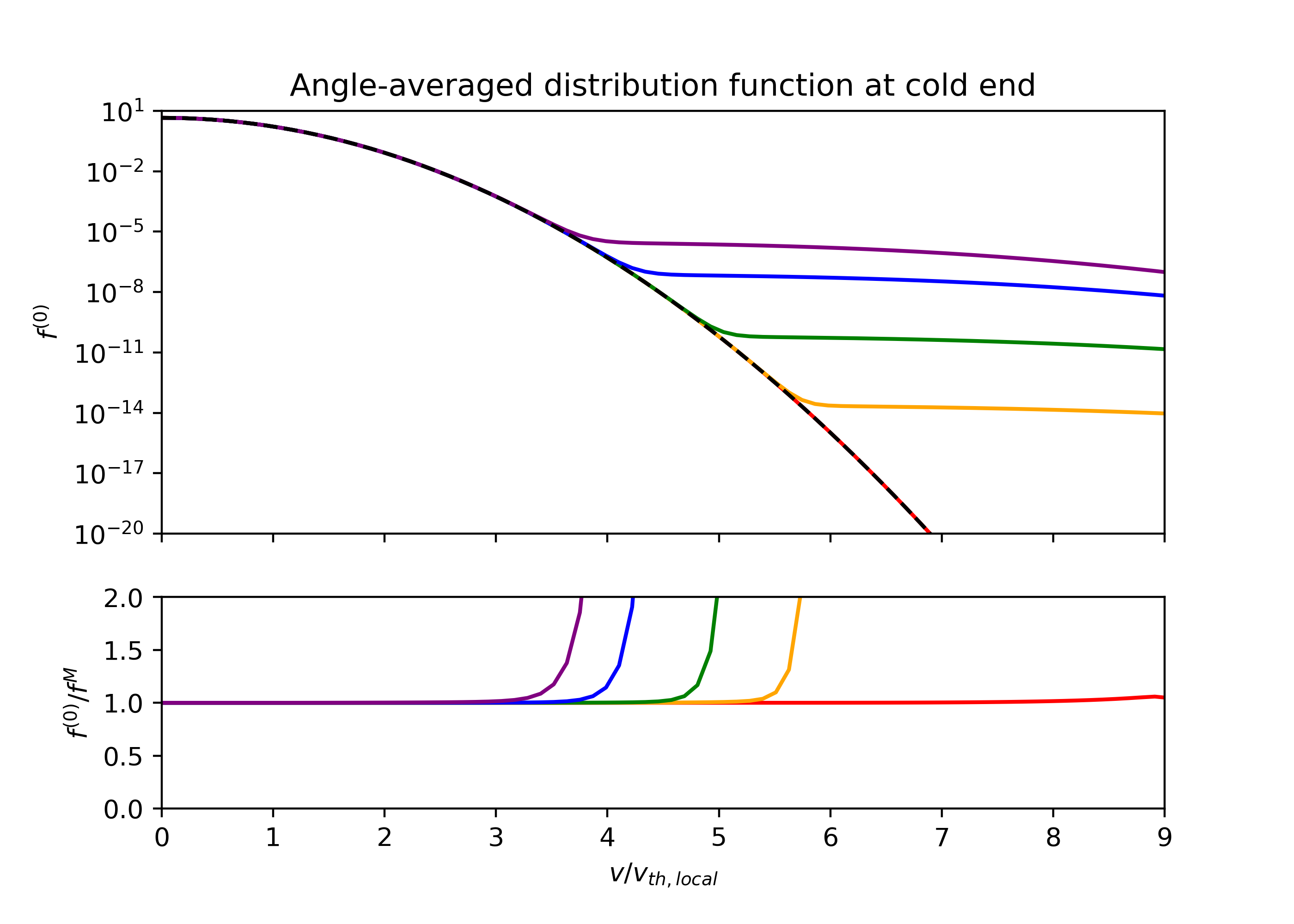}%
  }\\
  \subfloat[$f^{(1)}$ in hot end]{%
    \includegraphics[width=0.49\textwidth]{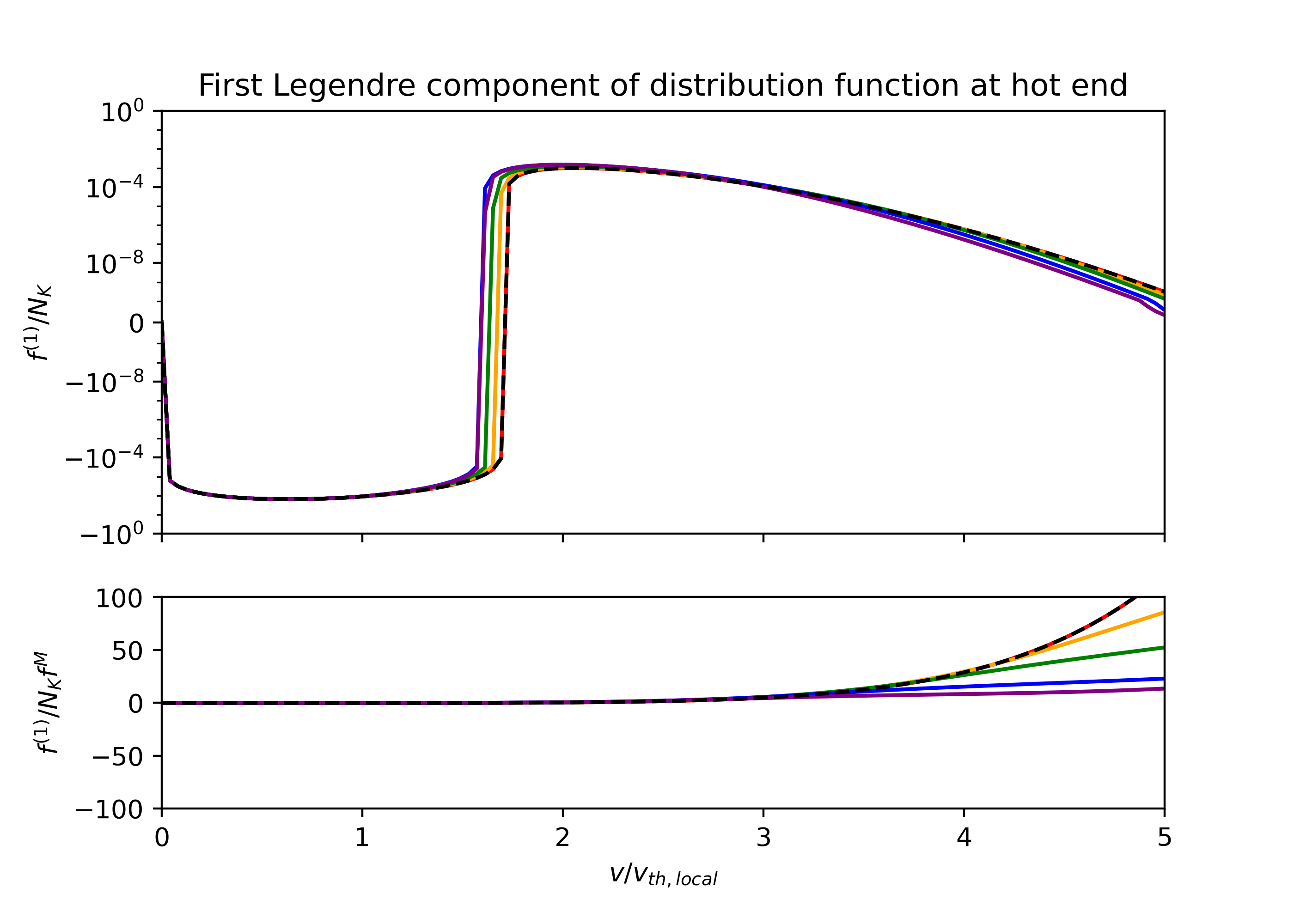}%
  }
  \caption{DF comparison at the hot end $z=-1.5L$ and cold end $z=+1.5L$ for various Knudsen numbers against local thermal speeds. Note the plots of $f^{(1)}$ have a symmetric log scale to include negative values. The smaller subplots show the ratio of each graph to the local Maxwellian with an appropriate scaling against Knudsen number included.}
\end{figure}
\begin{figure}
  \centering
  \ContinuedFloat
  \subfloat[$f^{(1)}$ in cold end]{%
    \includegraphics[width=0.49\textwidth]{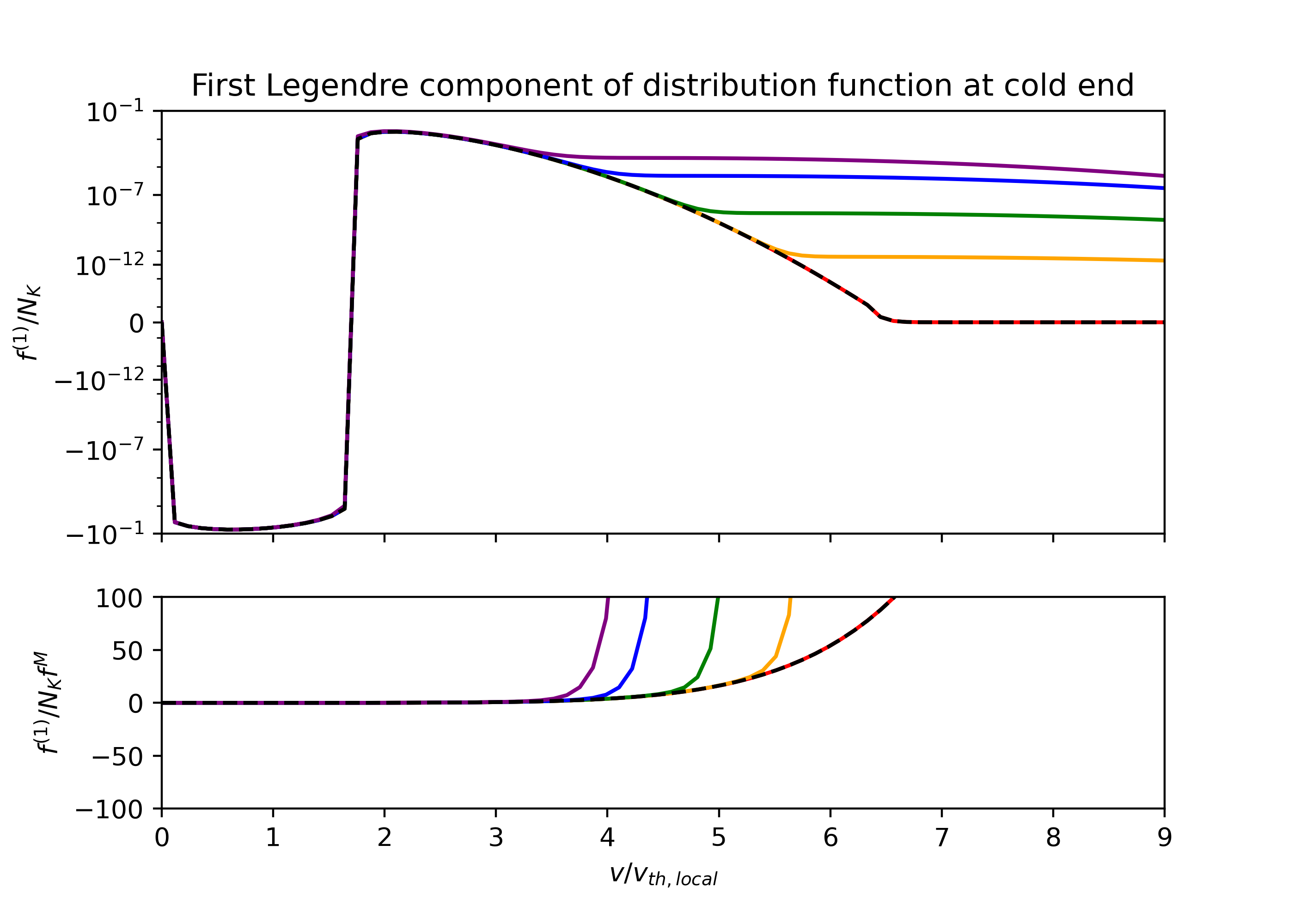}%
  }
  \caption{(continued)}
  \label{fig:single_species_DFs}
\end{figure}

Figure \ref{fig:single_species_DFs} shows the first two Legendre components $f^{(0)} = f^M + \delta f^{(0)}$ and $f^{(1)} = \delta f^{(1)}$ of the DF at the hot end and cold end for the single-species isochoric case. $f^{(0)}$ is the distribution function averaged over the velocity pitch angle $\theta$. For small Knudsen number $N_K = 1/1000$, the RKM result agrees with the local CE solution. For increasing Knudsen number, the tail of the DF in the hot end is further depleted and the tail of the DF in the cold end is further enhanced. This is expected physically at higher Knusden number as fast tail particles in the hot end may stream over the temperature gradient into the cold end, thereby reducing the tail population of the DF at the hot end and enhancing it in the cold end.

\begin{figure}
    \centering
    \includegraphics[width=0.49\textwidth]{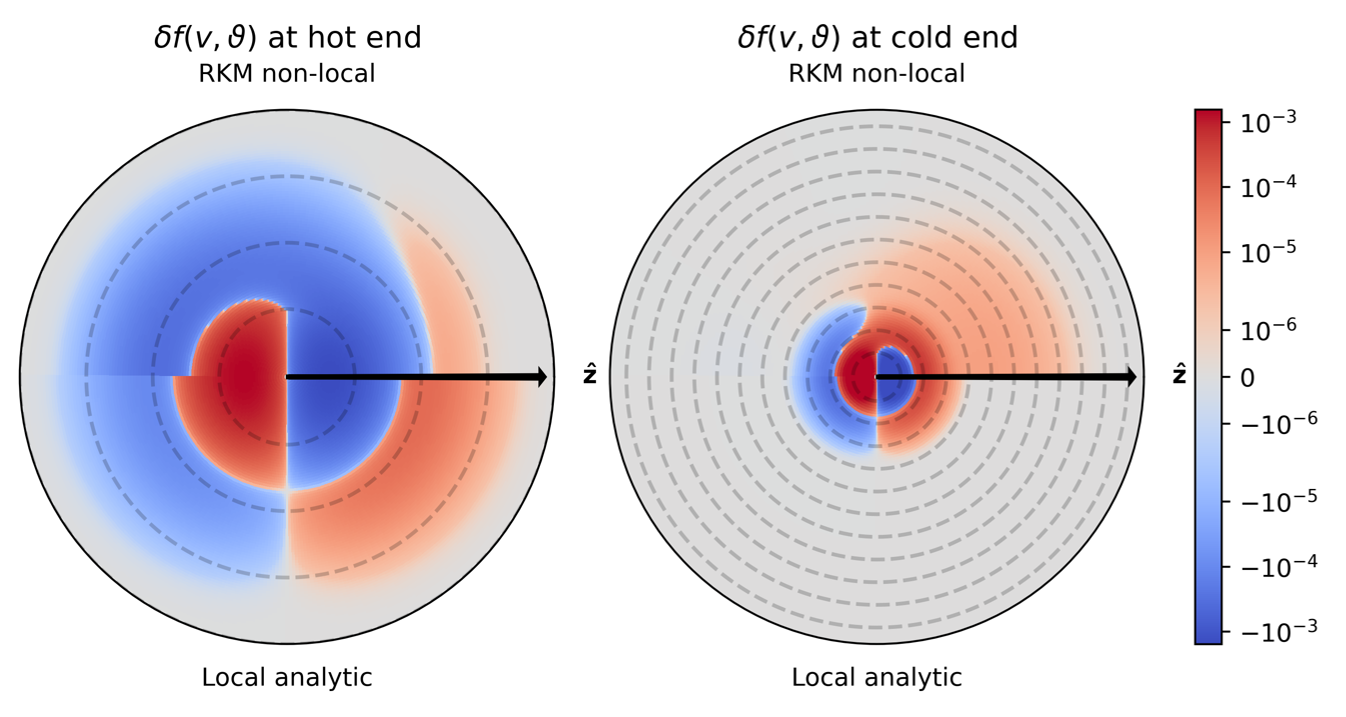}
    \caption{
        Deviation of the DF $\delta f (z,v,\vartheta)$ at the hot end $z=-1.5L$ (left) and cold end $z=+1.5L$ (right) in a non-local case with $N_K = 0.1$ (top half of each polar plot) and for the local analytic result (bottom half of each polar plot). Dashed circular contours indicating local thermal velocities are included. Note the color map uses a symmetric logarithmic scale with a linear section through zero.
    }
    \label{fig:polar_dfs}
\end{figure}
$\delta f (z,w,\vartheta)$ may be reconstructed via the Legendre series to visualize anisotropy in the DF as shown in figure \ref{fig:polar_dfs}. In the hot end, the tail of the DF is suppressed in the $\boldsymbol{\hat{z}}\cdot \boldsymbol{w} > 0 $ portion (noting $\boldsymbol{\hat{z}}\parallel -\boldsymbol{\nabla}T$) as is expected from faster tail particles with longer MFPs streaming out of the hot end, thereby depleting its tail population. In the cold end, the tail of the DF is enhanced in the $\boldsymbol{\hat{z}}\cdot \boldsymbol{w} > 0 $ portion due to suprathermal particles streaming into the cold end from the hot end. It may be noted that this tail enhancement appears particularly anisotropic with the portion of the tail with $\boldsymbol{\hat{z}}\cdot \boldsymbol{w} < 0 $ appearing mostly unchanged in the hot end, since the suprathermal particles from the hot end must have been travelling in the $\boldsymbol{\hat{z}}\cdot \boldsymbol{w} > 0 $ direction originally to enter the cold end, and they are subject to little collisional scattering in the cold end since $\nu_D$ is small in the tail. 

\subsection{Multi-species cases}

For multi-species cases, we define the Knudsen number in a similar manner as the single-species case, but instead using a maximum MFP $N_K = \lambda_{\text{max}} / L $ defined as $\lambda_{\text{max}} = \max_{\alpha,z} ( \lambda_{0,\alpha} (z) )$, with the MFP of a species $\alpha$ at spatial position $z$ accounting for collisions with all species via $ \lambda_{0,\alpha} = v_{\text{th},\alpha} / \sum_\beta \hat{\nu}_{\alpha\beta} $. A 50:50 DT mixture and a fully ionized CH\textsubscript{2} mixture are chosen as multi-species examples to investigate both weakly and moderately asymmetric mixtures. Both of these types of mixtures are relevant to ICF-related systems which often use DT fuel and plastic layers. A CH\textsubscript{2} mixture may be considered to be moderately asymmetric in mass, since the masses differ by an order of magnitude. Although the charge ratio is only $Z_C / Z_H = 6$, the charges enter the equations through the collision frequencies $\nu_{\alpha\beta} \propto Z_\alpha^2 Z_\beta^2$, therefore the charge asymmetry is effectively on an order of magnitude. Three cases are chosen for each mixture: a temperature gradient at uniform pressure and concentration, a pure pressure gradient at uniform temperature and concentration, and a pure concentration gradient at uniform temperature and pressure. These cases with isolated $\nabla T$, $\nabla p$, and $\nabla c$ are selected to investigate the sensitivity of the locality of heat fluxes to different driving gradients in temperature, pressure, and concentration. Similarly to the single-species case, the electric field has been set to zero to focus on the non-local effects from the aforementioned gradients. These six cases are shown in figures \ref{fig:DT_heat_flux} and \ref{fig:CH2_heat_flux} for DT and CH\textsubscript{2} respectively.

In a two-species ion mixture ignoring electric fields and electrons, the diffusive mass flux of the lighter species in the local limit may be written in the Landau form \cite{Landau1987Fluid,Kagan_2014}

\begin{equation}
    \begin{aligned}
        \rho_l \boldsymbol{V}_l = - \rho D \bigg( \boldsymbol{\nabla}c + k_p \boldsymbol{\nabla}\ln p + k_T \boldsymbol{\nabla}\ln T \bigg) \propto \boldsymbol{h},
    \end{aligned}
\end{equation}
where $D$ is the classical diffusion coefficient, and $k_p$ and $k_T$ are the baro-diffusion and thermo-diffusion coefficients respectively. It has been indicated that for a two-species mixture, the enthalpy flux $\boldsymbol{h}$ is directly proportional to the diffusive flux $\boldsymbol{V}_l$, therefore plots which show the enthalpy flux are directly related to the diffusive flux.

\begin{figure}
  \centering
  \subfloat[DT $\nabla T$]{%
    \includegraphics[width=0.49\textwidth]{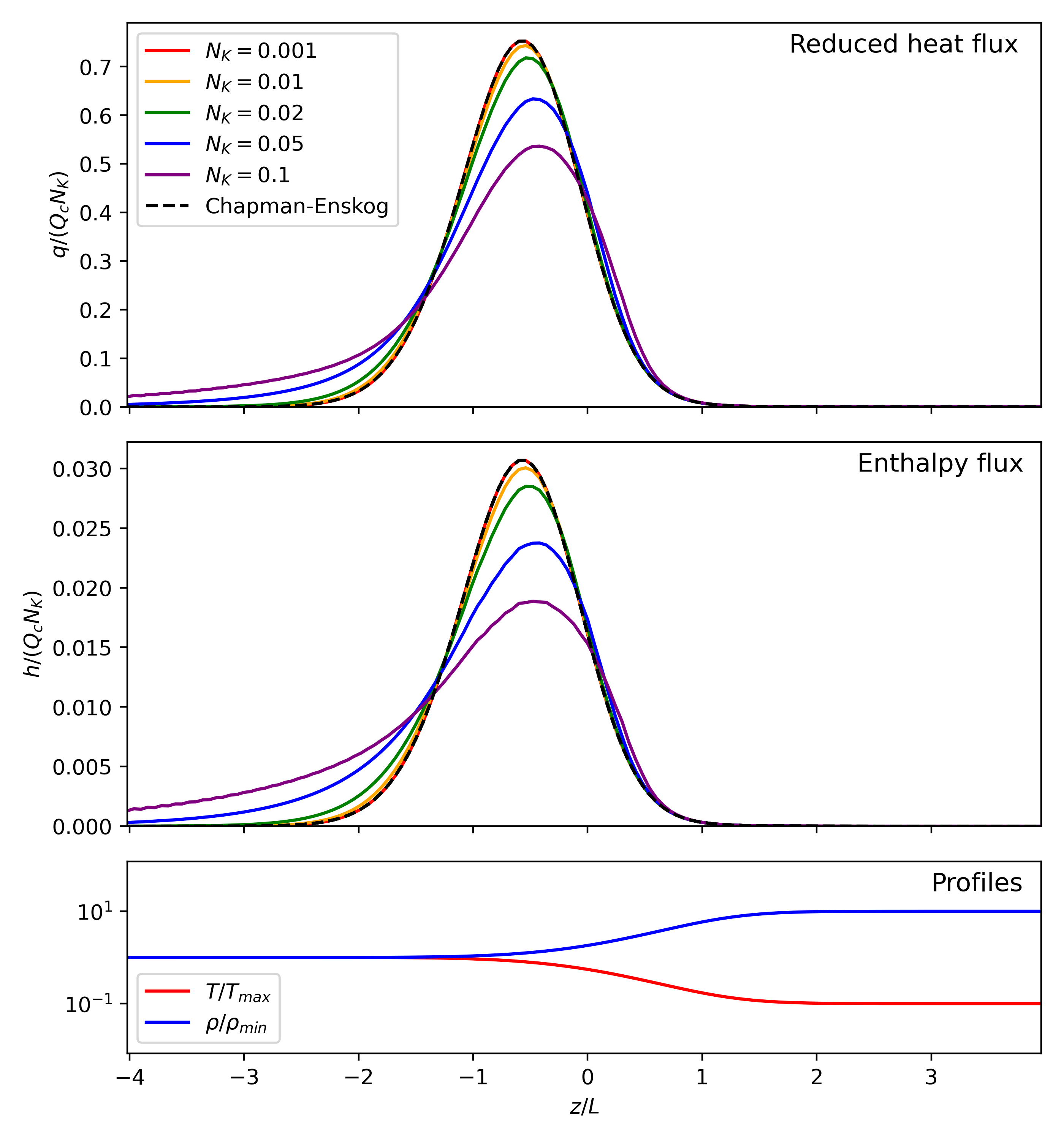}%
    \label{DT_grad_T}
  }\\
  \subfloat[DT $\nabla p$]{%
    \includegraphics[width=0.49\textwidth]{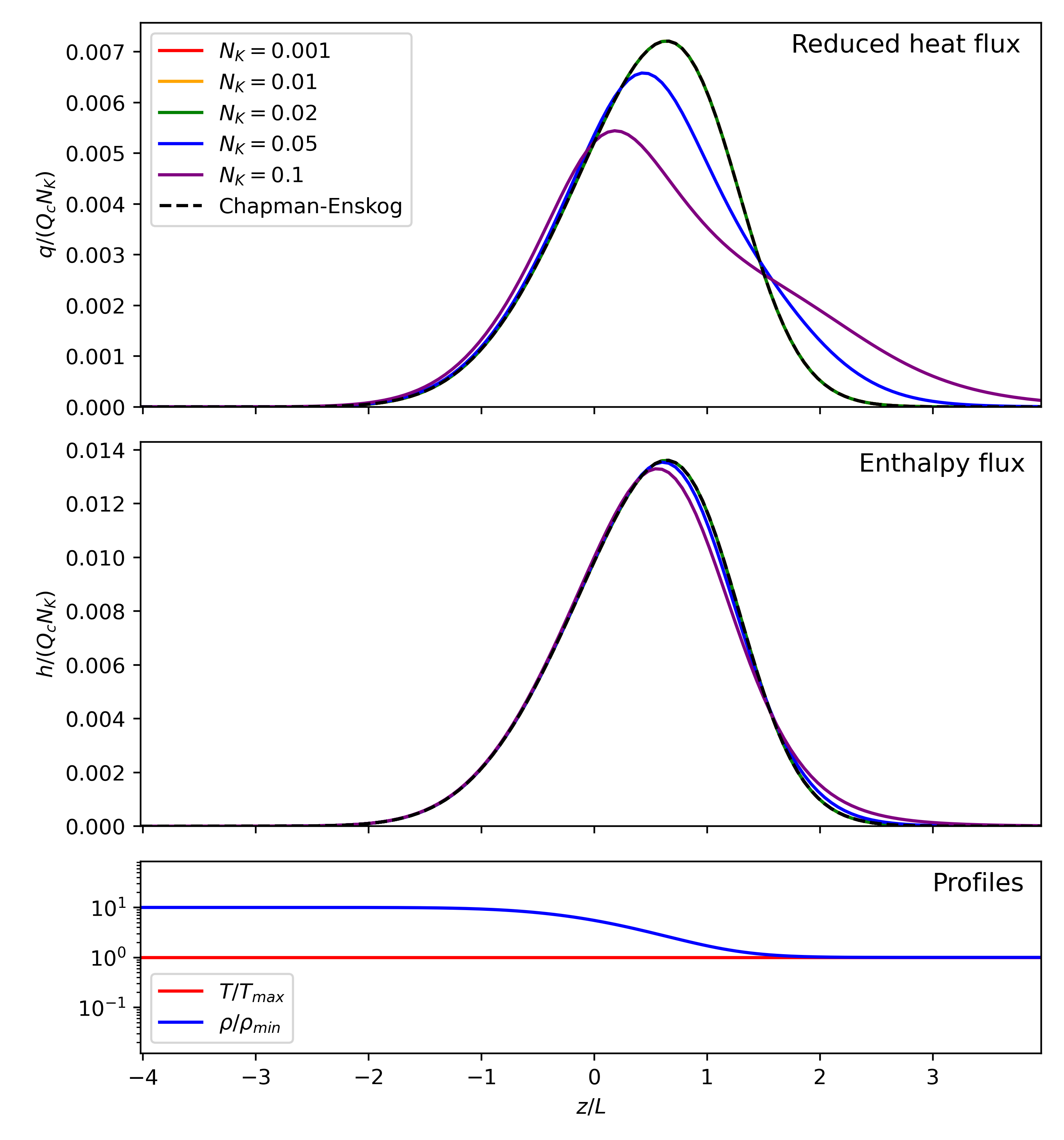}%
    \label{DT_grad_p}
  }\\
  \caption{Components of heat flux for a DT mixture subject to an isolated temperature gradient (a), isolated pressure gradient (b), and isolated concentration gradient (c). The shape of the relevant profiles are shown at the bottom of each subfigure, with undisplayed profiles being fixed constant at unity or concentrations at $0.5$. Results for $N_K \leq 0.01$ are not visible due to agreement with the local result.}
\end{figure}
\begin{figure}
  \centering
  \ContinuedFloat
  \subfloat[DT $\nabla c$]{%
    \includegraphics[width=0.49\textwidth]{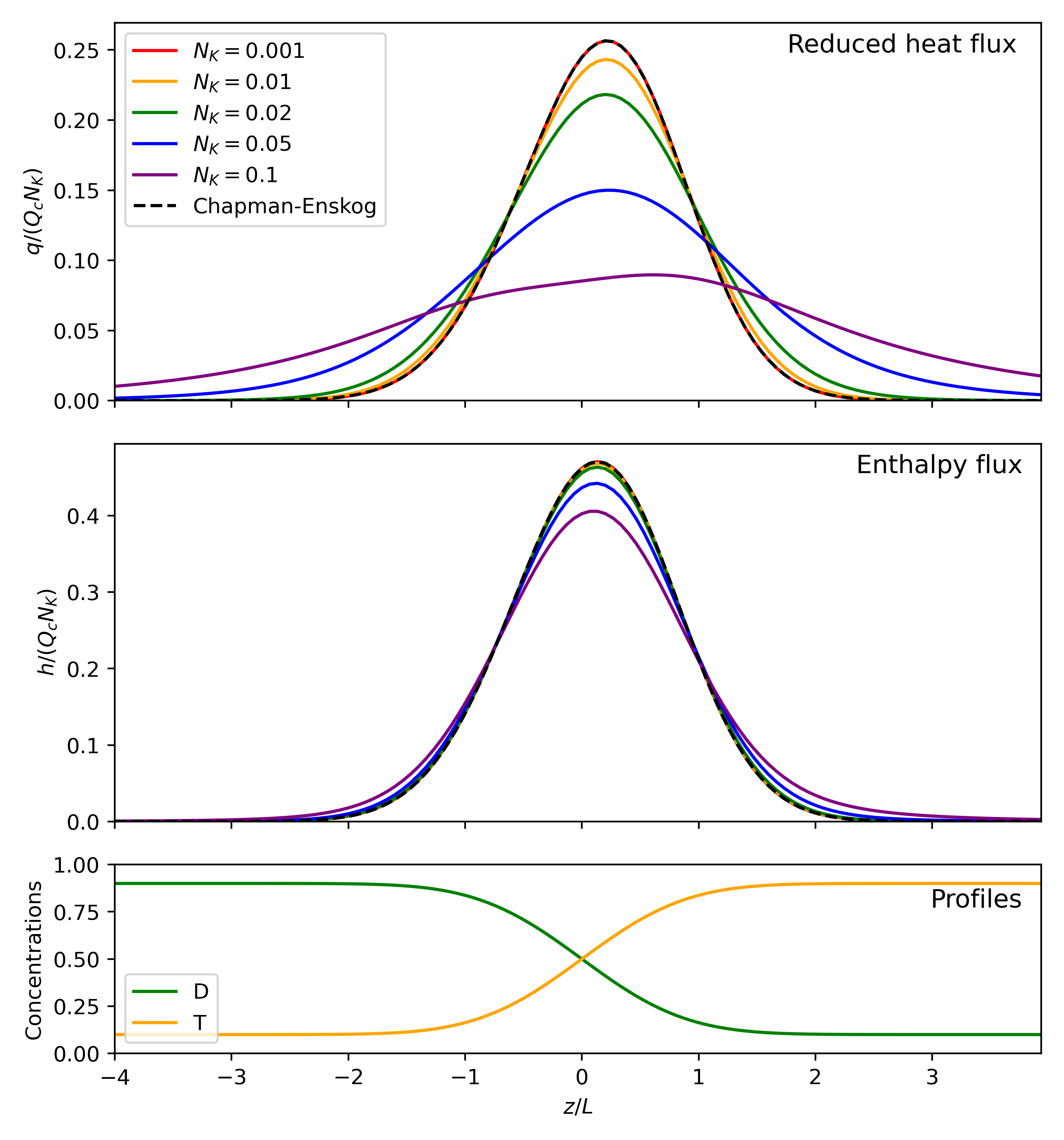}%
      \label{DT_grad_c}
  }
  \caption{(continued)}
  \label{fig:DT_heat_flux}
\end{figure}

\begin{figure}
  \centering
  \subfloat[CH\textsubscript{2} $\nabla T$]{%
    \includegraphics[width=0.49\textwidth]{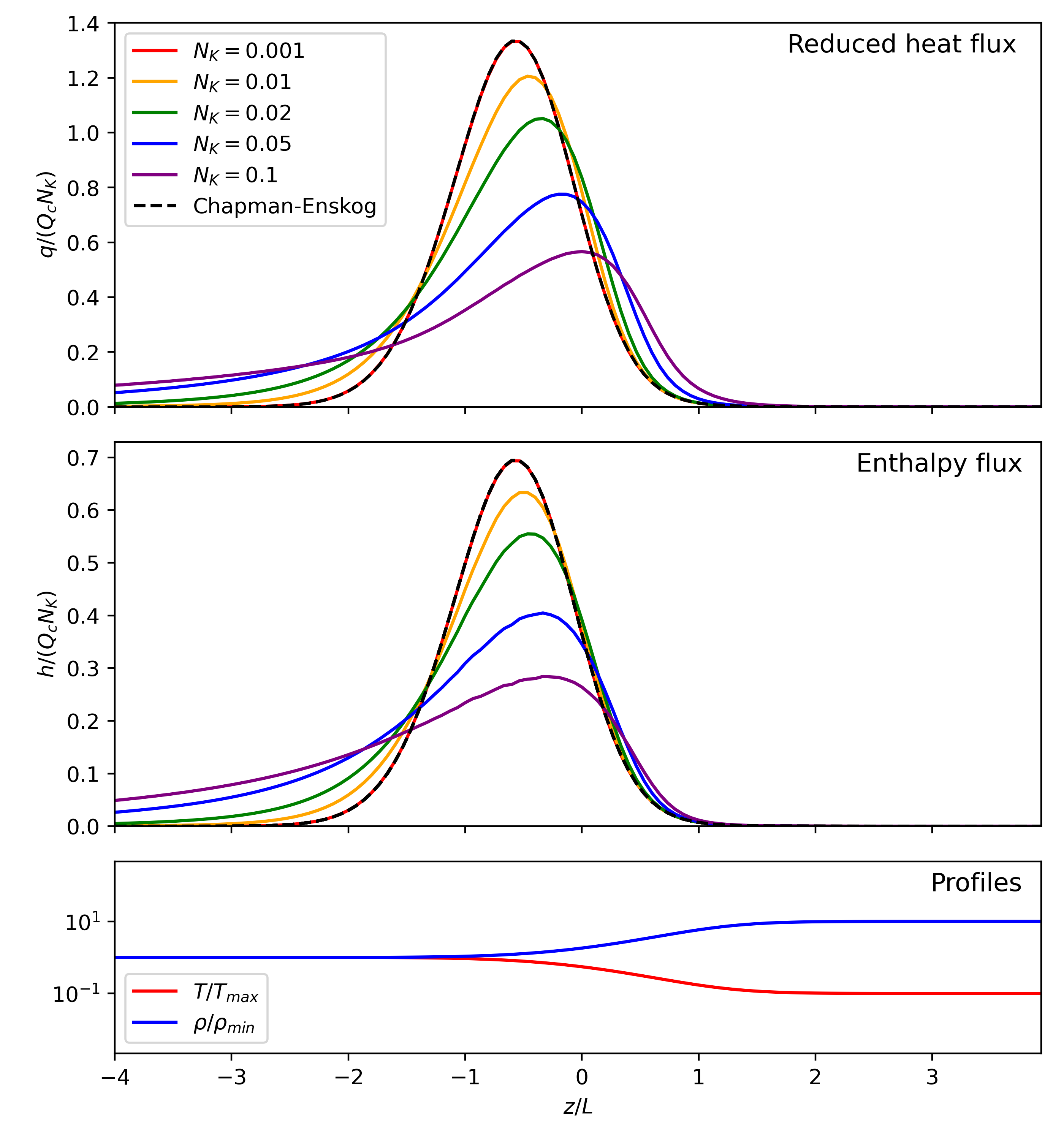}%
    \label{CH2_grad_T}
  }\\
  \subfloat[CH\textsubscript{2} $\nabla p$]{%
    \includegraphics[width=0.49\textwidth]{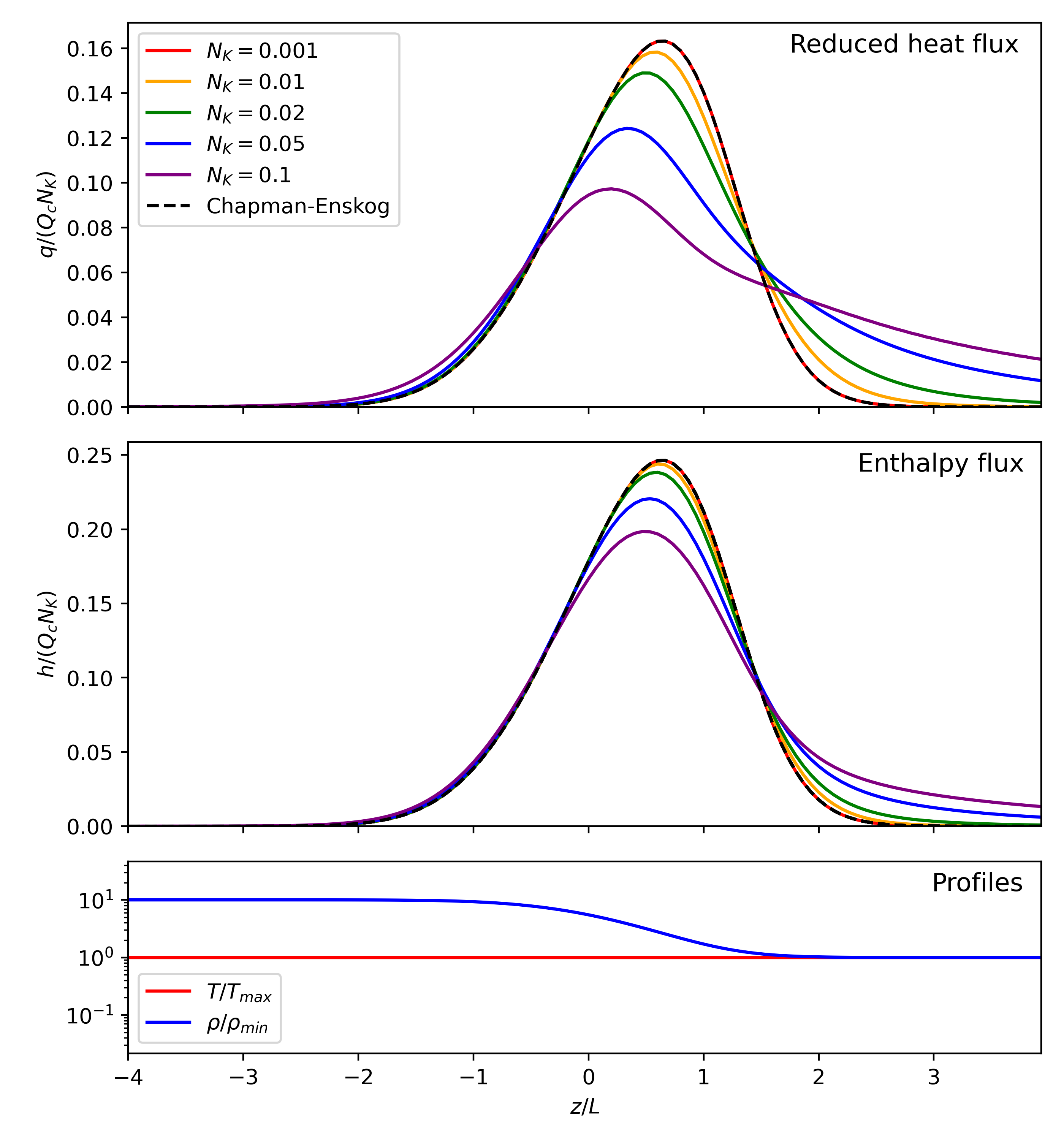}
    \label{CH2_grad_p}
  }\\
  \caption{Components of heat flux for a CH\textsubscript{2} mixture subject to an isolated temperature gradient (a), isolated pressure gradient (b), and isolated concentration gradient (c). The shape of the relevant profiles are shown at the bottom of each subfigure, with undisplayed pressure or temperature profiles being fixed constant at unity or concentrations such that $n_H/n_C = 2$.}
\end{figure}
\begin{figure}
  \centering
  \ContinuedFloat
  \subfloat[CH\textsubscript{2} $\nabla c$]{%
    \includegraphics[width=0.49\textwidth]{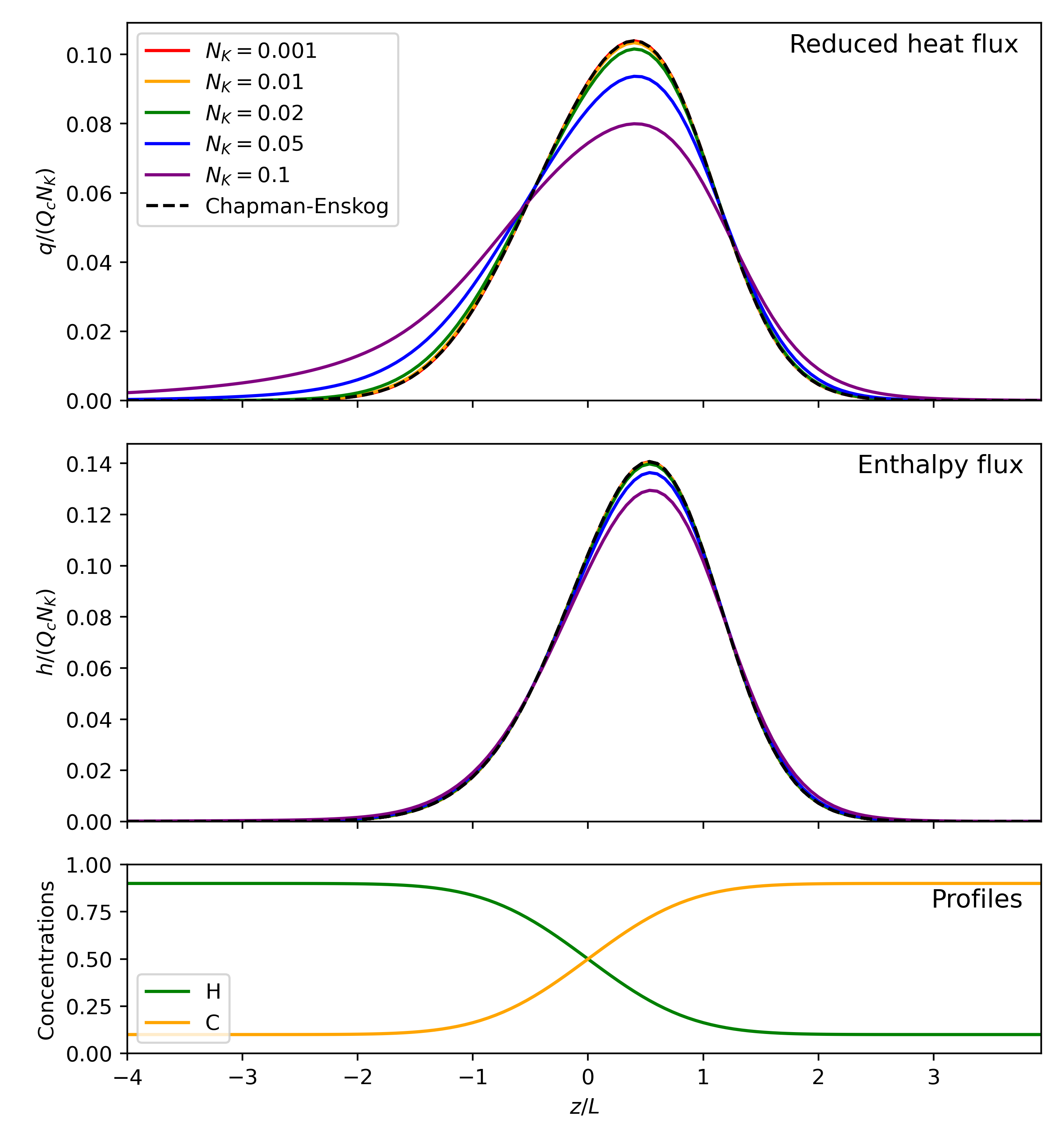}%
    \label{CH2_grad_c}
  }
  \caption{(continued)}
  \label{fig:CH2_heat_flux}
\end{figure}

All of the multi-species cases show the same basic qualitiative behaviour as the single-species cases; increasing Knudsen number inhibits the peak heat flux and enhances the heat flux in some regions away from the peak. However, the multi-species cases do demonstrate additional behaviour of interest. Firstly, non-local effects on the heat flux are demonstrated in the absence of a temperature gradient, induced instead due to sharp gradients in pressure or concentration. In many of the cases shown, non-local effects give rise to large heating pedestals in regions $\vert z / L \vert \gtrsim 2$ away from the boundary due to sharp gradients in $T$, $p$, or $c$. This is in dramatic contrast to the local heat flux in these regions, which is exponentially small for the chosen $\sim \tanh (z/L)$ profiles. In most cases this enhancement only appears in one direction to the boundary, for example the $\nabla p$ cases \ref{DT_grad_p} \ref{CH2_grad_p} where the non-local behaviour is only pronounced for $z/L>0$. In contrast, the DT $\nabla c$ \ref{DT_grad_c} case has significant non-local behaviour in both directions due to the near-symmetry, which is not present in any of the other cases. 

In the temperature gradient cases \ref{DT_grad_T} \ref{CH2_grad_T}, both components of the heat flux are similarly sensitive to non-local behaviour, and appear similar to the single-species case. Since the enthalpy flux is proportional to the diffusive flux, non-local behaviour in the former imply that diffusion can be sensitive to kinetic effects at similar Knudsen numbers at which heat transport is sensitive. In contrast, the DT $\nabla p$ case \ref{DT_grad_p} and both $\nabla c$ cases \ref{DT_grad_c} \ref{CH2_grad_c} have the reduced heat flux $q$ being significantly affected by non-local behaviour, whereas the enthalpy flux $h$ is quite insensitive to non-local behaviour. This leads to cases around the peak flux where $q$ is significantly suppressed and $h$ is not, so the latter may be increasingly important and dominant at the peak in non-local cases.

\subsection{Electron transport}

While the main focus of this paper is the ion heat transport, the newly developed approach can be applied to the electron case as well by treating electrons as an ion species with $Z_e = -1$ and $m_e \ll m_i$. Since all other non-local heat flux models available to date are for the electron heat flux only, we have run this calculation to be able to benchmark our method against an earlier, electron heat flux approach.

\subsubsection{Local comparison to Braginskii}

\begin{figure}[h!]
    \centering
    \includegraphics[width=0.49\textwidth]{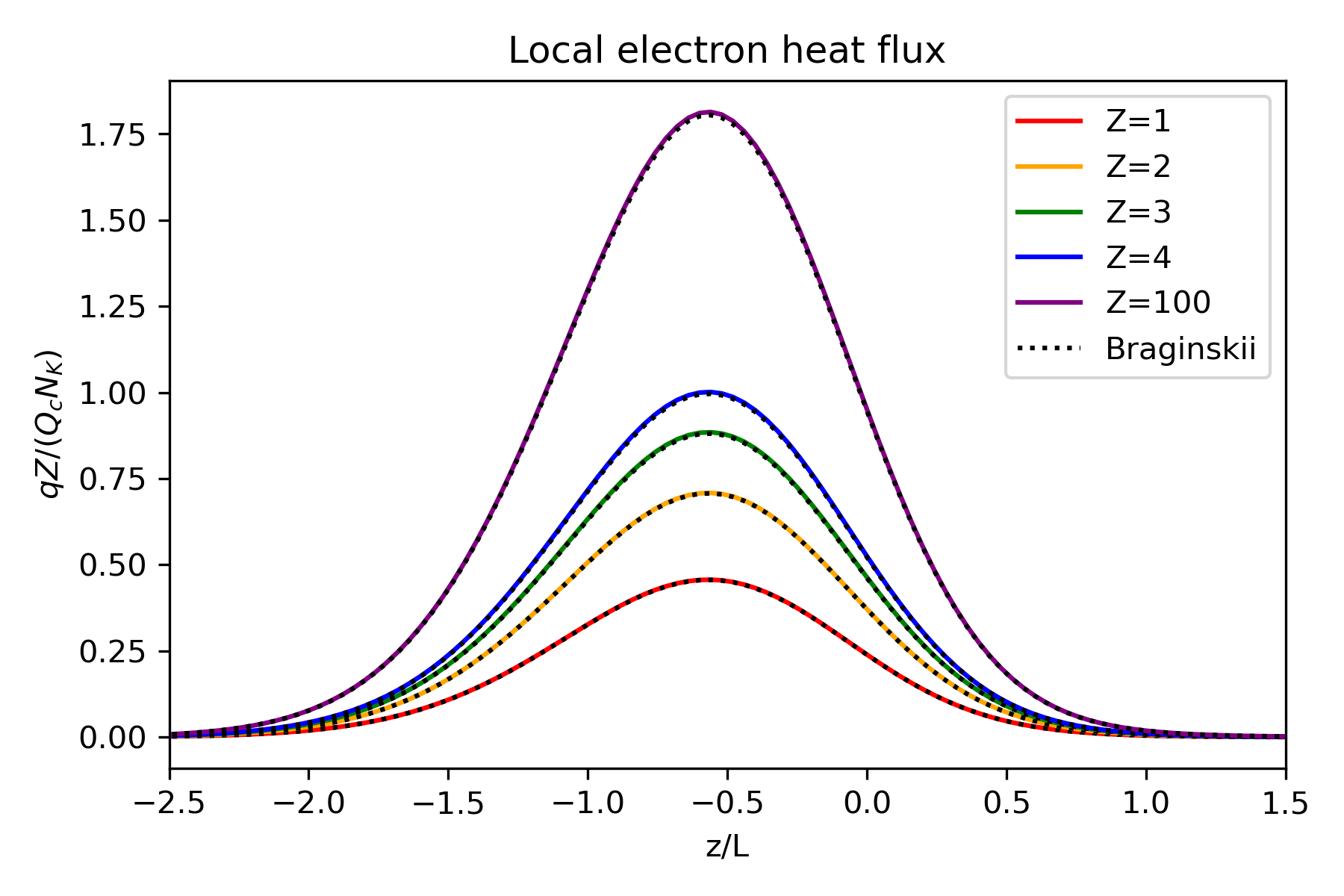}
    \caption{
        RKM results for the electron heat flux for various ionization numbers ($Z_i=1,2,3,4,100$) matched with Braginskii's results. Each dashed black line corresponds to the Z of the nearest colored RKM line. The RKM solver is run with two species, electron and ion, with $m_i / m_e = 1000$ and $Z_e = -1$. The temperature profile with Knudsen number $N_K=1/1000$ has a temperature drop by a factor of $10$, and the number densities are chosen such that the mixture is quasi-neutral and isobaric. Similarly to Braginskii, an electric field ensuring electron force balance using the local result for the thermal force is included.
    }
    \label{fig:local_electron_heat_flux}
\end{figure}

In the local limit, Braginskii's results for the electron heat flux may be reproduced for a quasi-neutral isobaric simple plasma with an electric field satisfying local electron force balance

\begin{equation}
    E = \frac{1}{en_e} (-\partial_z p_e + R_e ), \quad R^{\text{local}}_e = - B_e n_e \partial_z T_e
\end{equation}
where $R_e$ is the electron thermal force, with the local result including the electron thermal force coefficient $B_e$ appearing in (2.9) of \cite{Braginskii1965ReviewsOP}. The terms including flows, friction and viscosity, have been neglected. In the local limit, the RKM reproduces Braginskii's results successfully as shown in figure \ref{fig:local_electron_heat_flux}.

\subsubsection{Non-local comparison to SNB}

\begin{figure}[!htpb]
  \centering
  \subfloat[Full collision operator]{%
    \includegraphics[width=0.49\textwidth]{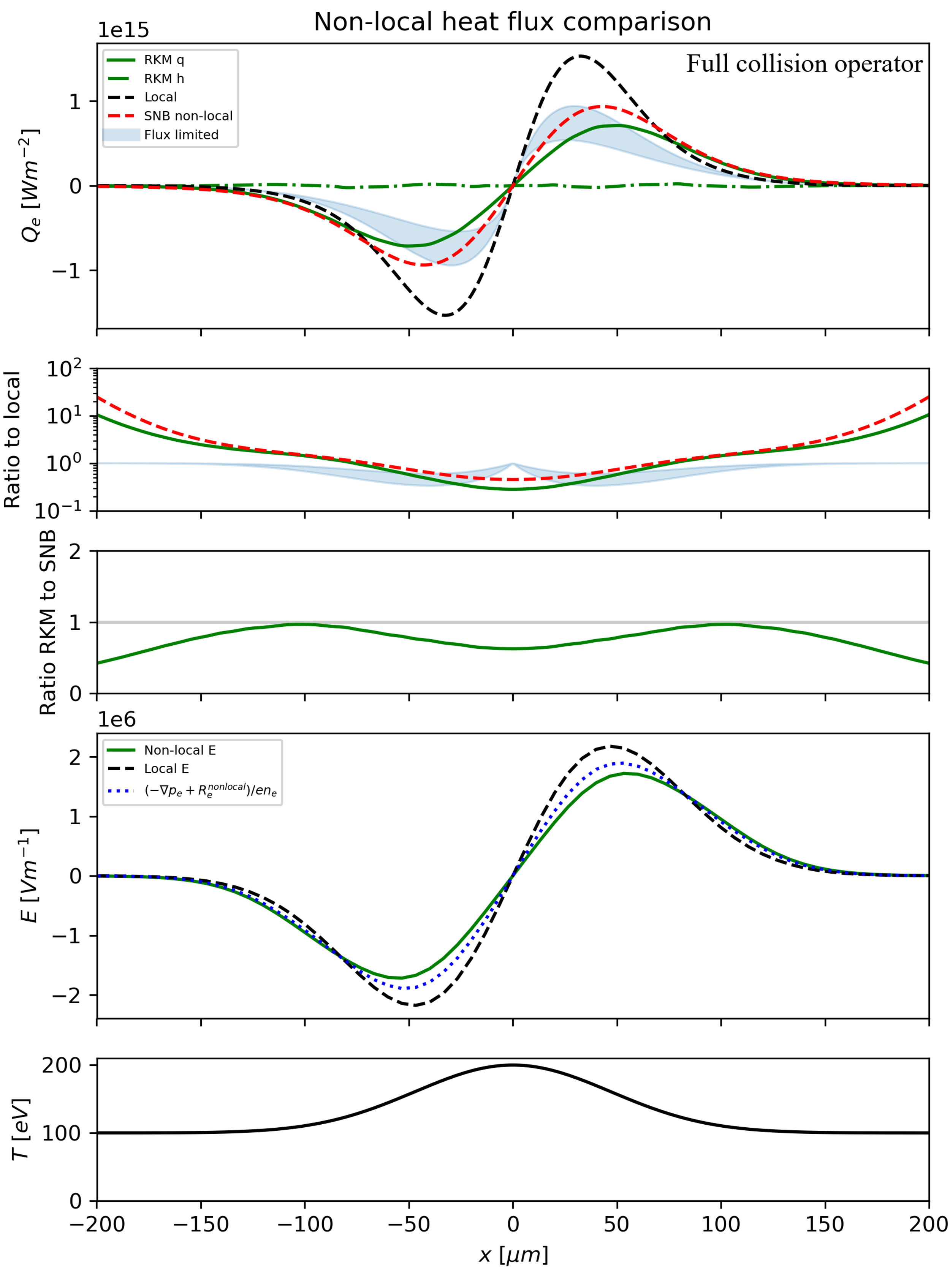}%
    \label{SNB_full_C_F}
  }\\
  \caption{Comparison of electron heat flux from the RKM solver, an SNB approach, and a flux limiter approach for an isochoric simple hydrogen plasma with a Gaussian temperature profile. (a) uses the full linearized Coulomb collision operator, and (b) uses the Coulomb collision operator with the field-particle part fixed at the local result from the Chapman-Enskog solution.}
\end{figure}
\begin{figure}
  \centering
  \ContinuedFloat
  \subfloat[Fixed field-particle part]{%
    \includegraphics[width=0.49\textwidth]{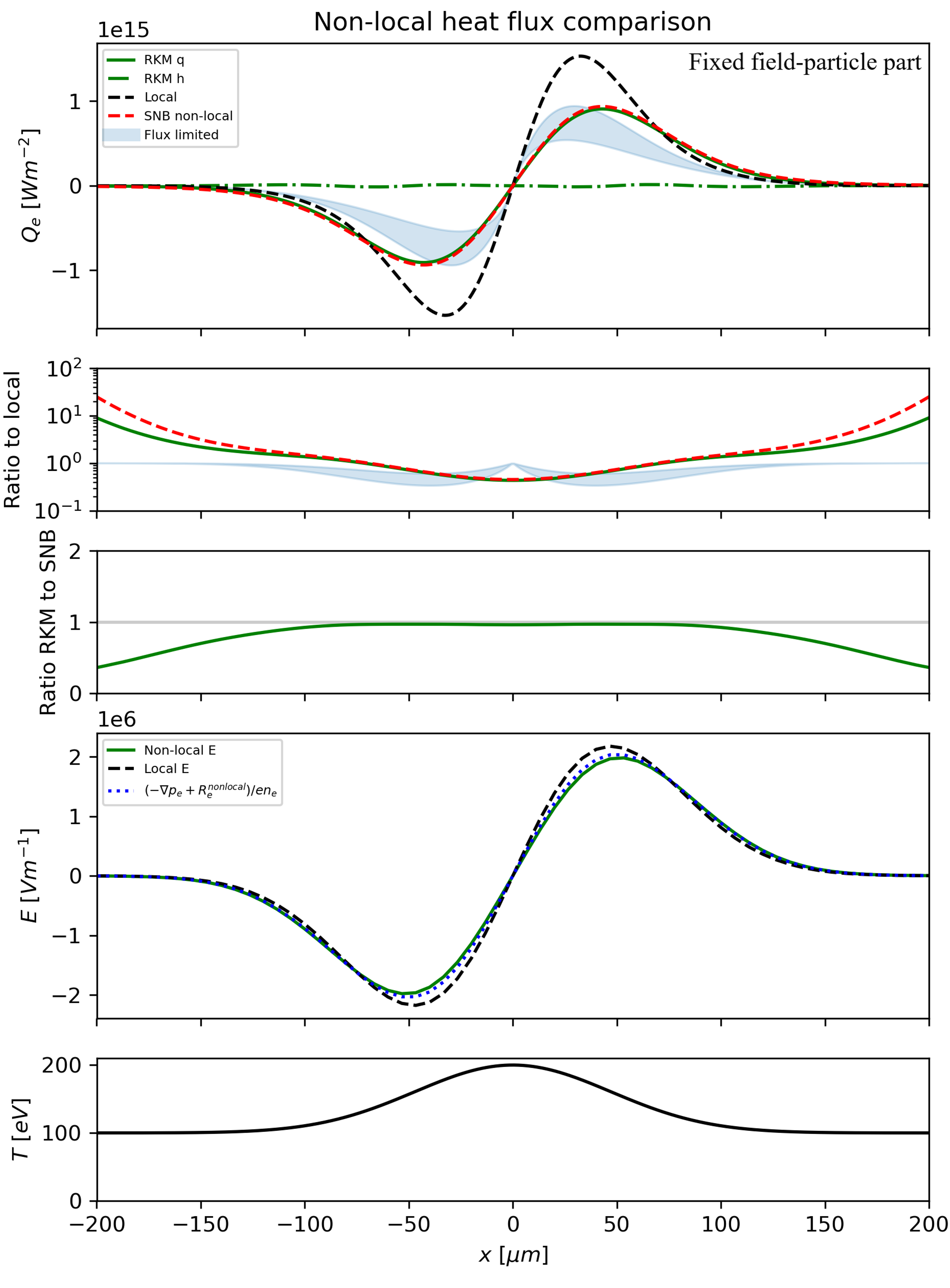}%
    \label{SNB_fixed_C_F}
  }
  \caption{(continued)}
  \label{fig:SNB_comparison}
\end{figure}

The RKM heat flux may be compared to the heat flux as predicted by other non-local transport models, namely the SNB approach and flux limiters. Figure \ref{fig:SNB_comparison} shows a comparison of the non-local electron heat flux from the RKM solver and SNB approach for an isochoric simple hydrogen plasma with a Gaussian temperature profile. For the chosen values $n_e = 10^{26}\ \text{m}^{-3}$, $T_{\text{max}} = 200\ \text{eV}$, $T_{\text{min}} = 100\ \text{eV}$, and $\ln \Lambda = 10$, the characteristic electron mean free path at the hottest temperature is $\lambda_0 = 6.16\ \mu \text{m}$. The SNB implementation used here follows the `iSNB' algorithm from Cao et al. \cite{10.1063/1.4928445}. The RKM and SNB hyperparameters are suitably chosen for convergence.

Braginskii's result for the electron conductive heat flux $q^T_e$ in the local limit chooses an electric field such that electron momentum conservation is satisfied, which includes the electron friction and thermal forces determined self-consistently from $\delta f_e$. In non-local cases, the local expressions for electron friction and thermal forces are no longer valid, therefore the correct electric field yielding zero current is not known before running the solver. Root-finding methods are applied to find the electric field in non-local cases such that the electron current, and therefore enthalpy flux, is zero everywhere.

In addition to SNB, a flux-limited heat flux is included for comparison. The flux limiter takes the simple harmonic-averaged form

\begin{equation}
    q_{\text{f.l.}} = \frac{q_{\text{Brag.}}}{1 + \vert q_{\text{Brag.}} / (\alpha q_{\text{f.s.}}) \vert  },
\end{equation}
where $q_{\text{f.l.}}$ is the flux-limited heat flux, $q_{\text{Brag.}}$ is the local Braginskii heat flux result, $q_{\text{f.s.}} = n_e T_{\text{min}} v_{\text{th},\text{min}} $ is a characteristic free-streaming heat flux for the system, and $\alpha$ is the flux-limiter parameter. Here, a range of values $0.05 \leq \alpha \leq 0.15$ are chosen. The flux-limited heat flux fails to capture enhanced heating in the colder region in comparison to the SNB and RKM approaches. 

The first case uses the full linearized Coulomb collison operator with test-particle and field-particle operator. The second case uses a modified Coulomb collision operator where the field-particle operator is fixed at its local result, i.e. $C \{ \delta f \} = C^T \{\delta f \} + C^F \{ \delta f^{\text{local}} \}$. This modification is proposed for improving computational performance while retaining some physical accuracy as the expensive integrals in the field-particle operator are not computed each timestep, yet agreement in the local limit is retained. As demonstrated, this approximation yields excellent agreement with SNB. The SNB approach is based on a simplified collision operator, with this particular implementation based on a Bhatnagar–Gross–Krook (BGK) type operator $C \{\delta f^{(1)} \} = - \nu_{ei}(w) \delta f^{(1)} $ with $ \nu_{ei}(w) \propto w^{-3}$. The improved agreement between the SNB and RKM approaches when the latter employs the fixed, rather than first-principle, field-particle operator suggests that the field-particle operator, which is non-local in velocity space, may play an important role in non-local behavior that is not captured in simpler collision models such as BGK that are local in velocity space.

\subsection{Convergence studies}

There are a few hyperparameters of the model which should be chosen for fast computational performance while converging to the physical solution. The two hyperparameters of most interest are the number of terms kept in the Legendre expansion $N_l$ and the minimum velocity cutoff $\xi_{\text{min}} = w_{\text{min}} / v_{\text{th}}$. 

The RKM solver is run for an isochoric single-species setup similar to the isochoric case in \ref{subsection_single_species} for a variety of values of $N_l$ with $N_K = 1/1000, 1/20, 1/10$ to investigate convergence in different cases of non-locality. The root-mean-square (RMS) deviation $\sigma_q = \frac{1}{Q_c N_K \vert D \vert} \sqrt{ \int_D dz\ (q - q_{\text{ref.}})^2 }$ is introduced to measure convergence of the heat flux, where $D$ is the spatial domain and $\vert D \vert$ is the length of the domain. The reference heat flux $q_{\text{ref.}}$ is chosen from a solution of the RKM solver with identical parameters as $q$ apart from the hyperparameter of focus.

\begin{figure}
    \centering
    \includegraphics[width=0.49\textwidth]{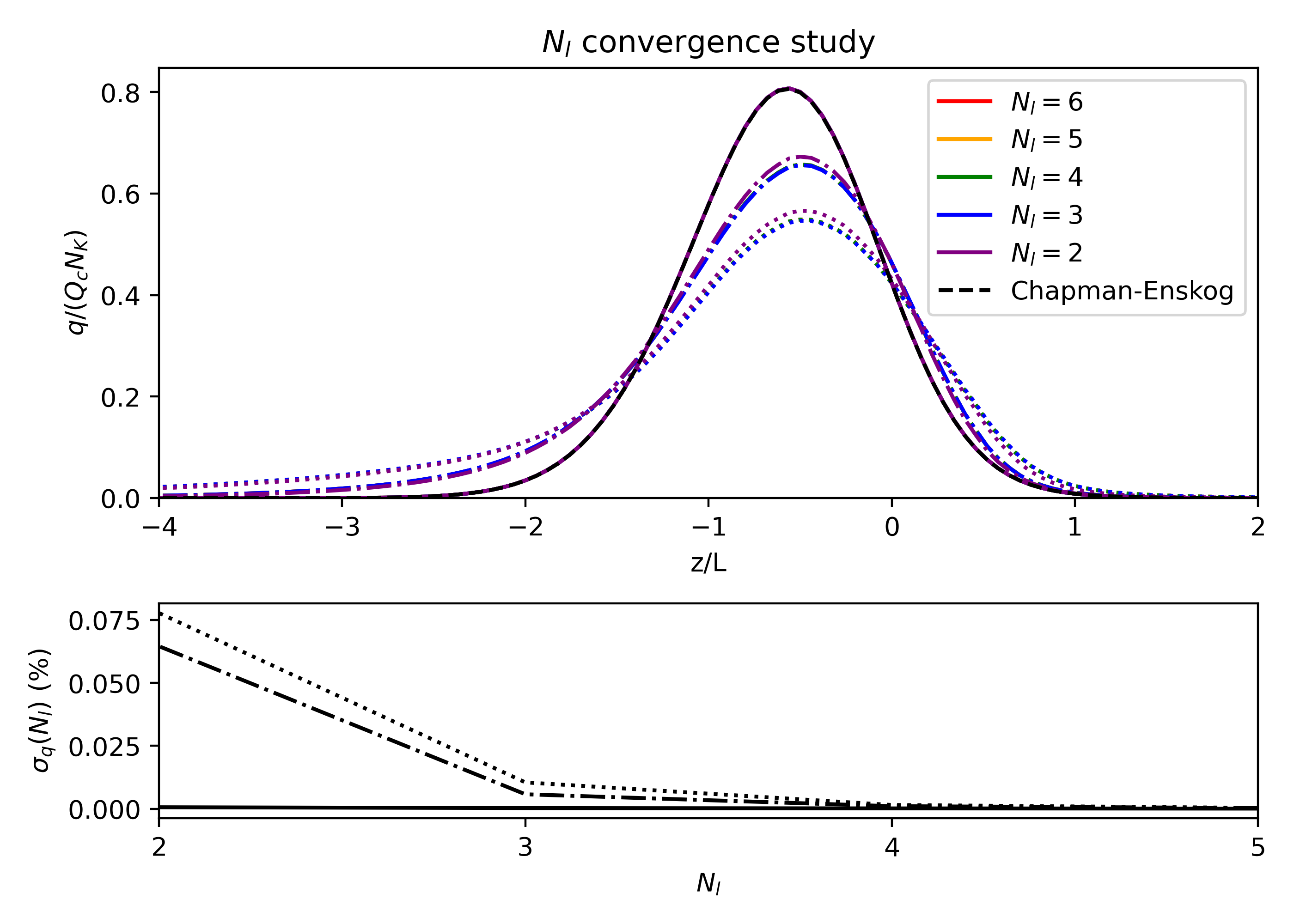}
    \caption{
        Convergence study for the number of Legendre modes included in the expansion $\delta f \approx \sum_{l=0}^{N_l-1} P_l(\mu) \delta f^{(l)}$. The solver is run identically to the isochoric single species example \ref{single_species_fluxes} with $\xi_{\text{min}} = 1$, with the local limit results ($N_K=1/1000$) shown in dashed lines and two significantly non-local cases ($N_K=1/20$ and $N_K = 1/10$) shown in dotted-dashed lines and dotted lines respectively. In the local cases, the RKM solutions are difficult to see due to good agreement with the CE solution. The RMS deviation is also shown. Here, $q_{\text{ref.}}$ uses $N_l = 6$ results since more Legendre components included in the expansion should improve the accuracy of the solver.
    }\label{fig:N_l_convergence}
\end{figure}

Figure \ref{fig:N_l_convergence} shows the convergence study for $N_l$. Good agreement with the local analytic solution is practically identical in all cases as expected, since the local solution for $\delta f^{(l)}$ is only nonzero for $l=1$. In non-local cases, good convergence is demonstrated even for $N_l = 4$. Larger $N_l$ increases convergence time, in particular scaling with $\mathcal{O}(N_l)$. The default value $N_l = 6$ is chosen.

For the low velocity cutoff, a value around $\xi_{\text{min}} \approx 1$ should be taken since a smaller value includes the bulk in the RKM solver domain, which gives rise to large computational cost and numerical instabilities, and a larger value does not account for parts of the tail that have significant contribution to the heat flux. In the region $0.9 \leq \xi_{\text{min}} \leq 1.1$, there is a variation of $\sigma_q < 0.1 \% $ for $N_K \leq 1/10$, therefore the heat flux is mostly insensitive to this hyperparameter for a choice which appropriately defines the tail. This further validates the approach of solving only for the tail of the distribution beyond some cutoff since the RKM and local Chapman-Enskog solutions match smoothly at the low velocity boundary. The default value $\xi_{\text{min}} = 1$ is chosen. 


\subsection{Temperature self-consistency}

The RKM solver takes input number density and temperature profiles and solves for the tail of the DF of each species. In extreme non-local cases, a sufficiently non-Maxwellian tail may result in the density and temperature moments of the resulting DF to significantly differ from the input profiles, leaving an inconsistent solution. Since the temperature is a higher moment than the density, the former is more sensitive to deviation due to non-local tail behaviour and more likely to be inconsistent. The RKM temperature is defined as

\begin{equation}
    T^{\text{RKM}}_\alpha(z) = \frac{m_\alpha}{3n_\alpha}\int d^3\boldsymbol{w}\ w^2 f_\alpha = \frac{4\pi m_\alpha}{3n_\alpha}\int_0^\infty dw\ w^4 f^{(0)}_\alpha.
\end{equation}

\begin{figure}
  \centering
  \subfloat[Single species]{%
    \includegraphics[width=0.49\textwidth]{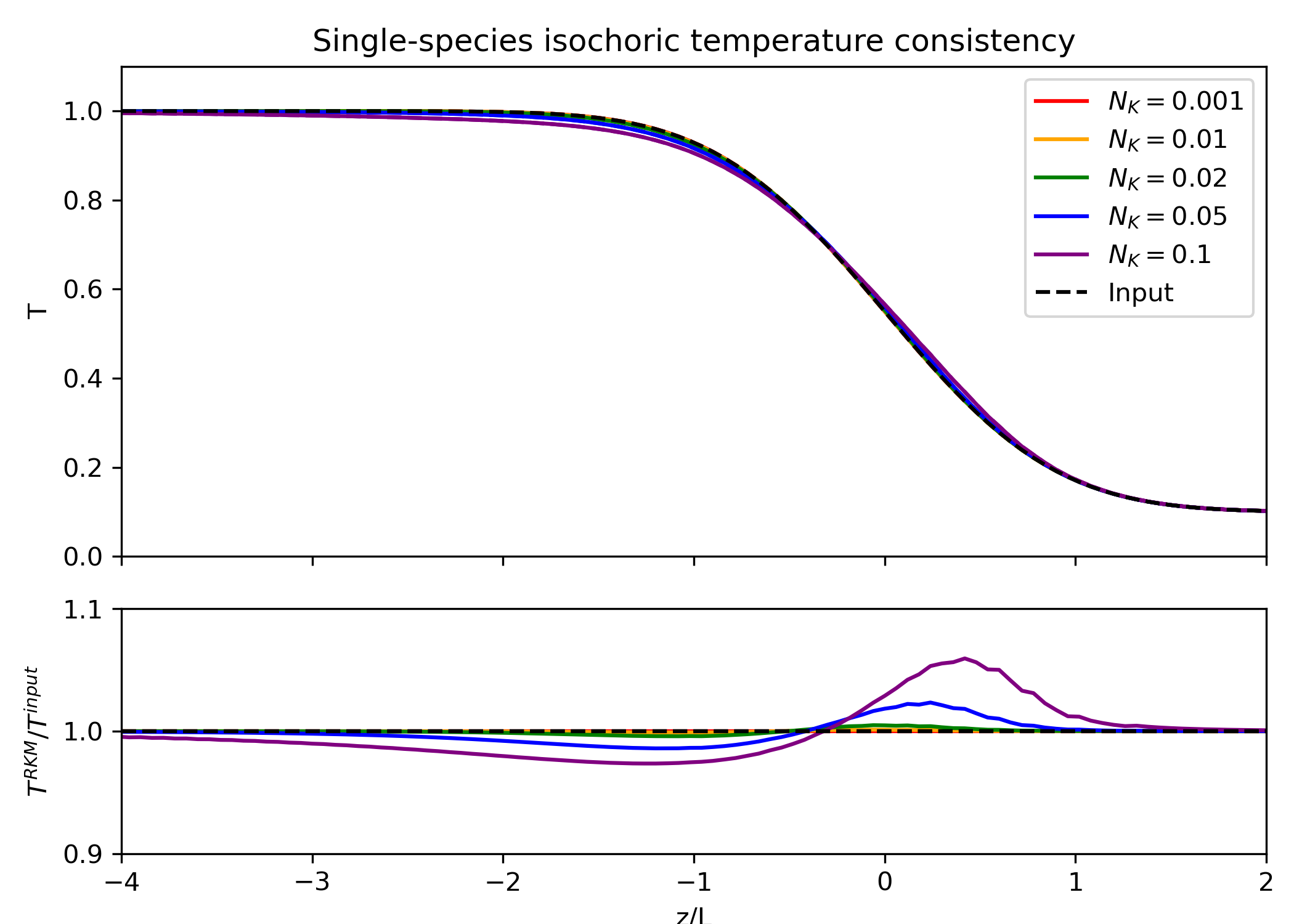}\label{fig:single_species_temperature_consistency}%
  }\\
  \subfloat[DT]{%
    \includegraphics[width=0.49\textwidth]{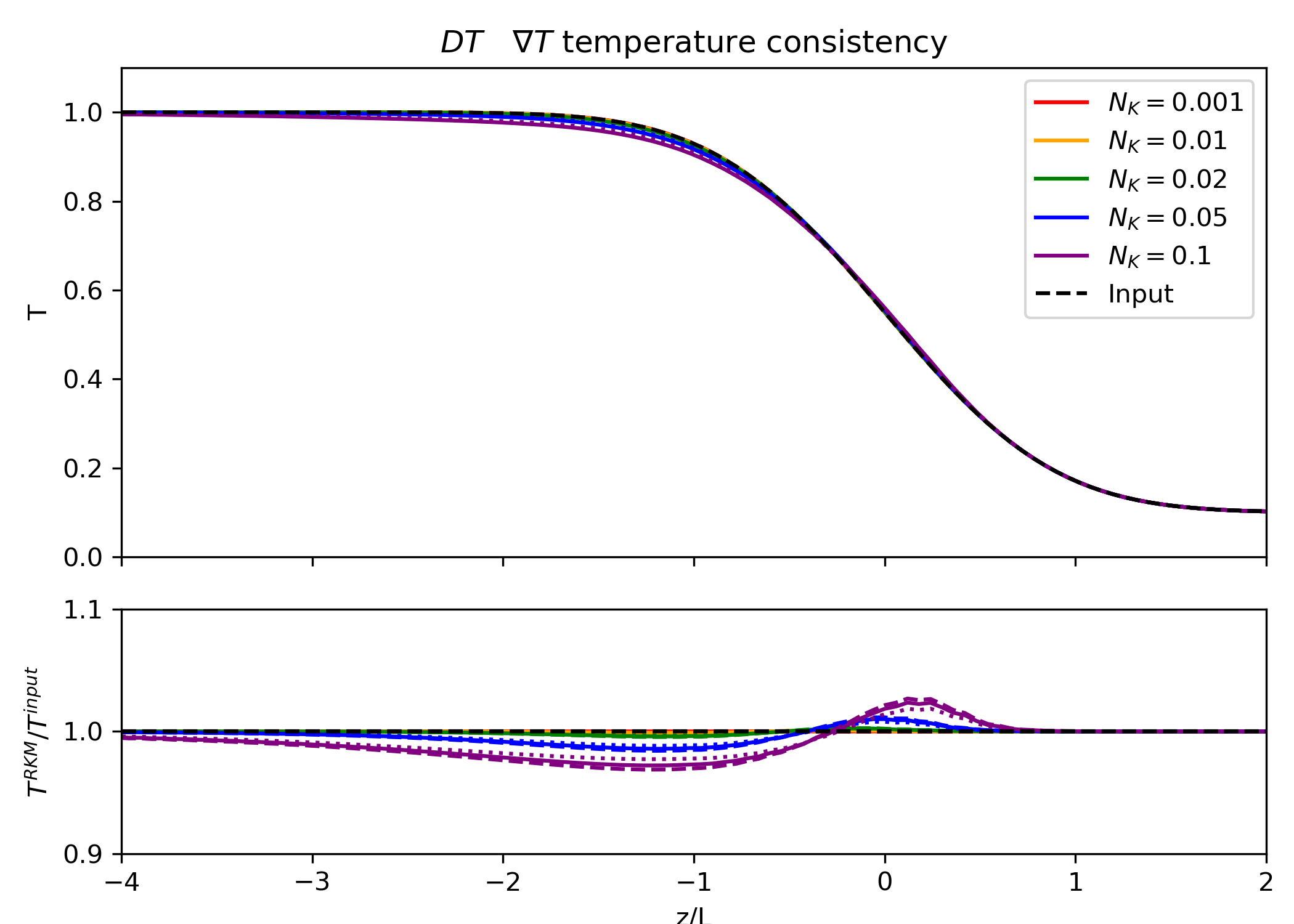}
    \label{fig:DT_temperature_consistency}
  }\\
  \subfloat[CH\textsubscript{2}]{%
    \includegraphics[width=0.49\textwidth]{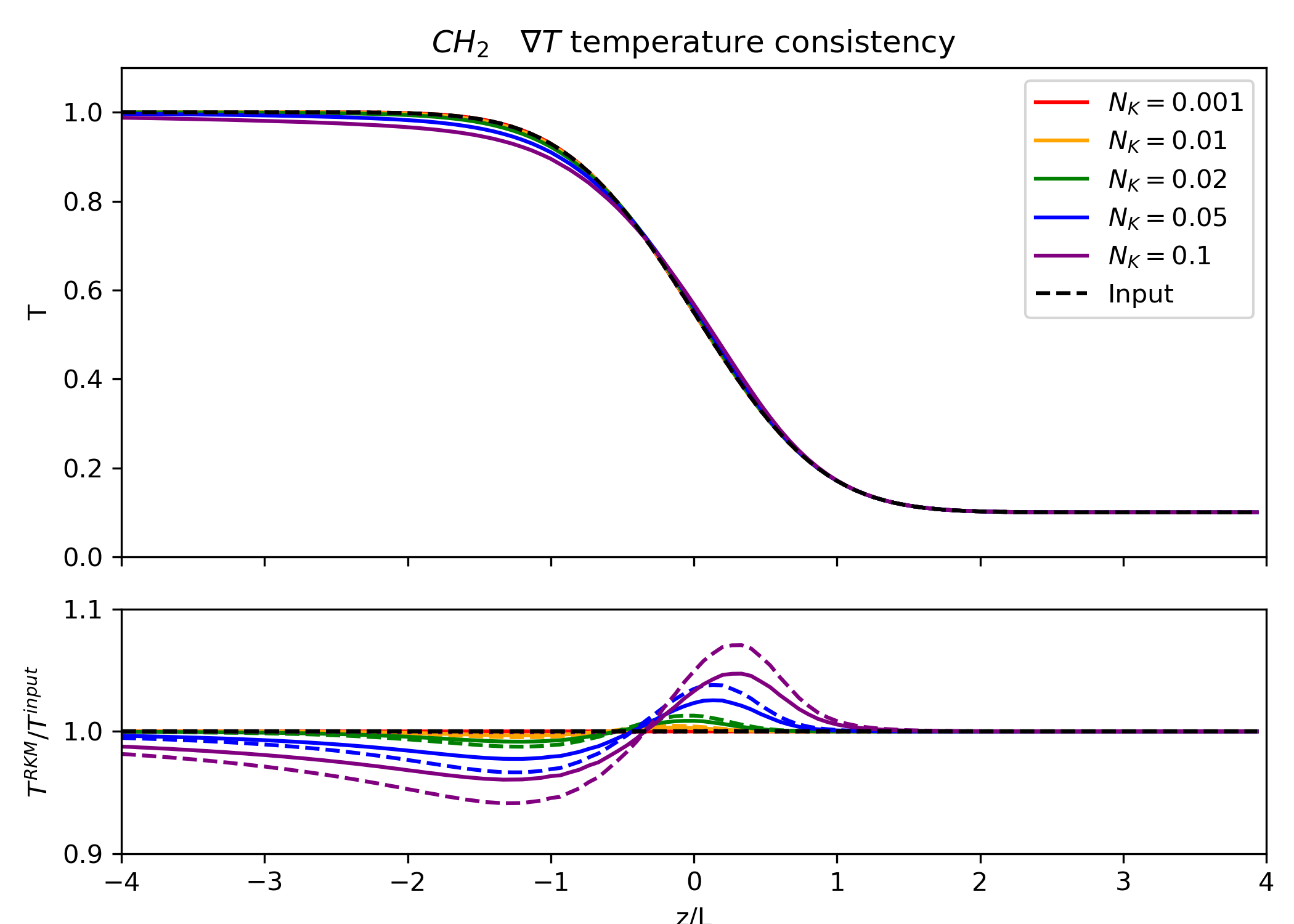}
    \label{fig:CH2_temperature_consistency}
  }\\
  \caption{
        RKM temperature for the single-species isochoric case (a), DT $\nabla T$ case (b), and CH\textsubscript{2} $\nabla T$ case (c). The upper subplots show the RKM integrated temperature, and the lower subplots shows the ratio of the RKM temperature to the input temperature profile. For the multi-species cases, the lighter species ($D$ or $H$) temperature is shown in dashed lines, the heavier species ($T$ or $C$) temperature is shown in dotted lines, and the total ion temperature $T^{\text{RKM}}_{\text{total}} = \sum_\alpha n_\alpha T^{\text{RKM}}_\alpha / \sum_\alpha n_\alpha $ is shown in solid lines.
    }
  \label{fig:temperature_consistency}
\end{figure}

Figure \ref{fig:temperature_consistency} shows the RKM temperature for the previously discussed single-species isochoric case, DT $\nabla T$ case, and CH\textsubscript{2} $\nabla T$ case. In all three cases, the local cases are in good agreement with the local result, with inconsistency developing with increasing Knudsen number. Even for the extreme cases of $N_K = 0.1$, the total temperature is consistent to $\lesssim 5 \% $ deviation. In the multi-species cases, it is the temperature of the lighter species that is more inconsistent with the input profile. 

\subsection{Smallness of field-particle operator}

\begin{figure}
  \centering
  \subfloat[Hot end]{%
    \includegraphics[width=0.49\textwidth]{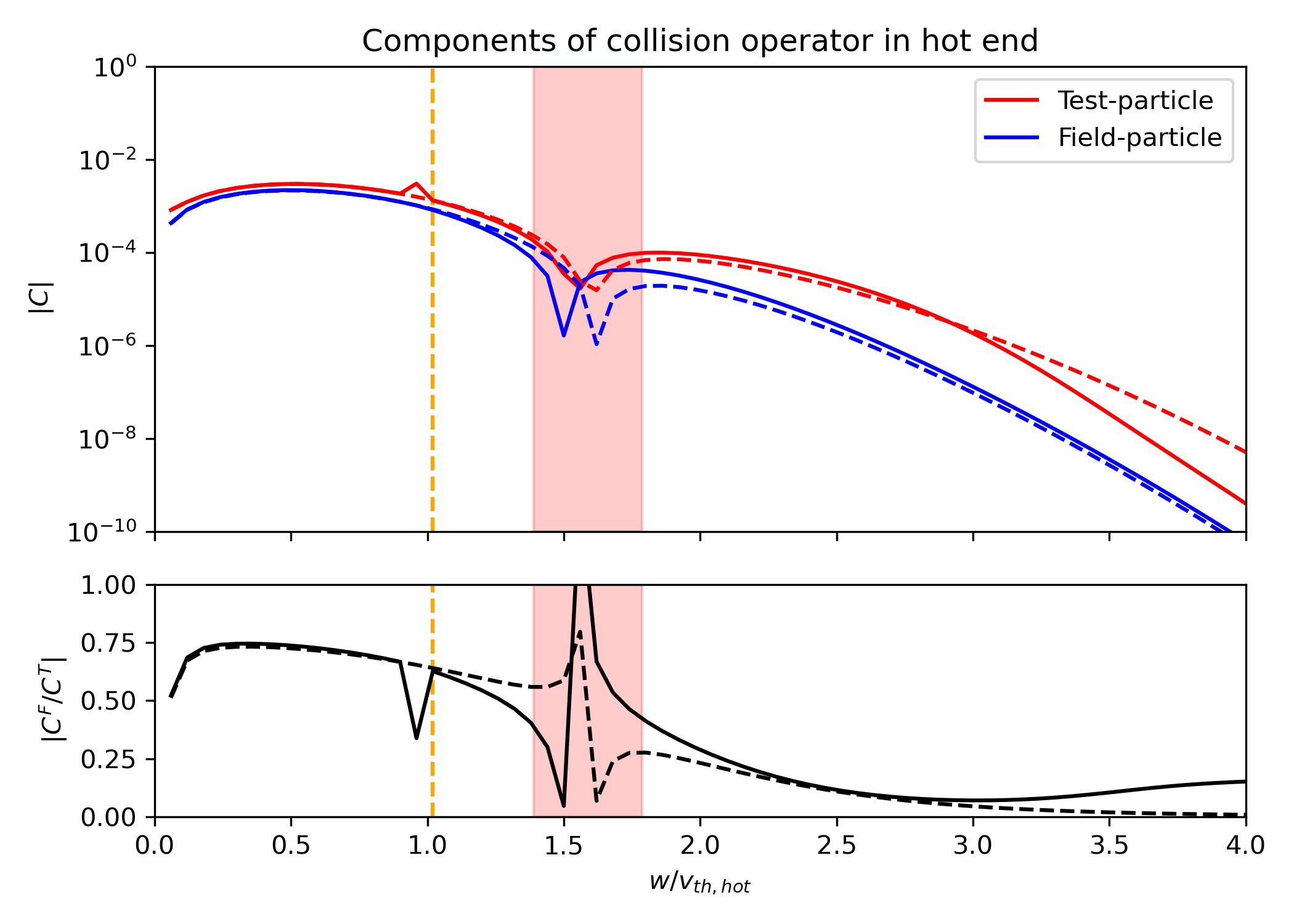}%
    \label{operator_comparison_hot}
  }\\
  \subfloat[Cold end]{%
    \includegraphics[width=0.49\textwidth]{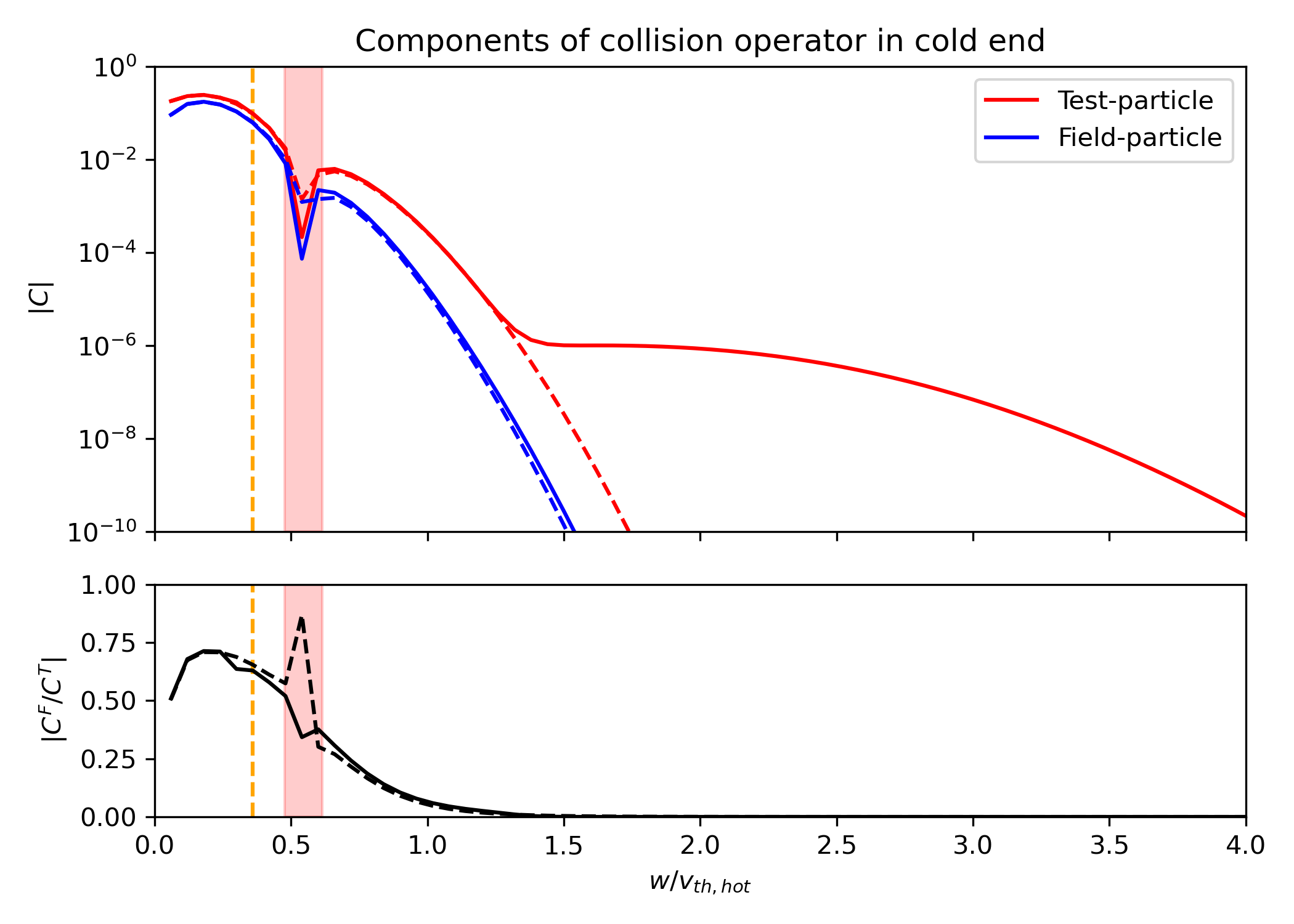}%
    \label{operator_comparison_cold}
  }\\
  \caption{Comparison of $l=1$ components of the test-particle operator and field-particle operator at the hot end (a) and cold end (b) from the single-species isochoric example. The solid lines show results for the $N_K=0.1$ results, and the dashed lines show the local analytic results. The lower subplots show the ratio of field-particle operator to test-particle operator.}
  \label{fig:operator_comparison}
\end{figure}

Figure \ref{fig:operator_comparison} shows a comparison of the $l=1$ components of the test-particle operator and field-particle operator at the hot end ($z/L = -1.5$) and cold end ($z/L = +1.5$) from the single-species isochoric example in \ref{subsection_single_species} for the non-local $N_K=0.1$ case against local analytic results. The vertical orange dashed lines indicates the local low velocity cutoff, chosen to be the local thermal velocity $w_{\text{min}} = v_{\text{th}}$. The solution for the tail above the low velocity cutoff $w \geq w_{\text{min}}$ is determined by the RKM solver. The solution below the low velocity cutoff in the bulk is fixed as the local Chapman-Enskog solution $\delta f( w < w_{\text{min}} ) = \delta f^{\text{local}} (w)$, therefore the test-particle operator is also identical for the RKM and local results away from the cutoff. In contrast, the RKM and analytic solutions of the field-particle operator may differ in this region due to the velocity integrals that sample the solution above the low velocity cutoff. This deviation is small due to the small population of the tail compared to the bulk, therefore the local analytic result should still hold as a good approximation of the bulk.

At the velocity grid point just below the low velocity cutoff there is a spike in the test-particle operator compared to the local solution. This originates from the $ w^{-2} \partial_w ( w^4 \nu_\parallel \partial_w \delta f)$ term, suggesting that the numerical scheme as implemented does not constrain continuity of the second velocity derivative of the DF continuous at the cutoff. This numerical artifact lies outside the domain of the RKM solver, and continuity of the second or higher derivatives of the DF are not required, therefore is not a point of concern.

The red highlight indicates the regions around $w / v_{\text{th},\text{local}} = \sqrt{5/2} \approx 1.6 $ where the local analytic result $\delta f_{\text{local}}^{(1)} \propto (w^2 / v_{\text{th}}^2 - 5/2)$ changes sign. The sudden spike of the ratio in this region is therefore not of concern since both the operators are passing through zero. Apart from these red regions, the field-particle operator remains small in comparison to the test-particle operator. The Jacobian of the field-particle operator is then a small difference between the approximate and true Jacobian, which explains why the approach of approximating the Jacobian of the collision operator in the tail without the non-sparse elements of the field-particle operator described in \ref{numerical_implementation} works numerically.

\section{Discussion}\label{section_conclusion}

A reduced kinetic method has been presented and applied to investigate non-local multi-species ion heat transport for the first time. The RKM solves for the tail of distribution functions according to the Vlasov-Fokker-Planck equation for a classical, unmagnetized multi-species plasma in a 1D2V geometry. In the single ion species case, the non-local ion  heat flux show qualitatively similar trends to its well explored electron counterpart. On the other hand, novel non-local features are recovered for the multi-species ion heat transport due to additional thermodynamic drives such as concentration gradients and diffusion of different species in the centre-of-momentum frame.

Results for DFs and heat fluxes in local cases reproduce the analytic results of the Chapman-Enskog expansion \cite{chapman_cowling,Ferziger_book,kagan2016transport} for multi-species ion heat flux and Braginskii's \cite{Braginskii1965ReviewsOP} results for ion and electron heat flux in a simple plasma. Non-local electron heat flux results agree well with an SNB implementation, particularly where the field-particle operator is fixed at the local solution rather than implemented in a first principle fashion. Expected qualitative behaviour of non-local effects are demonstrated, in particular suppression of the peak heat flux and enhancement of heat flux away from regions of steep gradient. In multi-species cases, non-local effects are demonstrated from gradients in temperature, concentration, and pressure, with some cases giving rise to large heat fluxes far away from the boundary where the local result is exponentially small. The multi-species heat flux has been decomposed into two components, the reduced heat flux and enthalpy flux, where the former is the term employed in mainline hydrodynamic codes and the latter is an inherently multi-species feature due to relative dynamics of the species in the center-of-momentum frame. As demonstrated here for the first time, in cases with pressure or concentration gradients, the enthalpy flux is particularly insensitve to non-local effects, thus becoming increasingly important in comparison to the reduced heat flux for high Knudsen numbers. The RKM has been applied here to describe non-local heat flux, but since the approach solves for the DF of each ion species, it may have future utility in investigating other kinetic effects such as reactivity reduction in inertial confinement fusion experiments \cite{McDevitt_2,PhysRevLett.115.105002,10.1063/1.4921130}. Future work relating to the heat flux may investigate additional effects such as magnetic fields and the implementation of a RKM prescription to non-local heat flux in a simple 1D hydrodynamics code.

\begin{strip}

\appendix

\section{Kinetic equation for deviation of distribution function from Maxwellian}
\label{appendix_kinetic_equation_for_deviation}

The kinetic equation may be transformed to the centre-of-momentum frame by transforming the velocity to the peculiar velocity $\boldsymbol{w} = \boldsymbol{v} - \boldsymbol{u}(\boldsymbol{t},\boldsymbol{x})$, leaving

\begin{equation}
    \begin{aligned}
        &(\partial_t - (\partial_t \boldsymbol{u}) \cdot \boldsymbol{\nabla_w} + (\boldsymbol{w} + \boldsymbol{u})\cdot (\boldsymbol{\nabla} - (\boldsymbol{\nabla u})\cdot \boldsymbol{\nabla_w}) + \boldsymbol{F}_\alpha \cdot \boldsymbol{\nabla_w}) f_\alpha = \sum_\beta C_{\alpha\beta}\{f_\alpha,f_\beta\}.
    \end{aligned}
\end{equation}
 
The DFs are then decomposed via $f_\alpha = f^M_\alpha + \delta f_\alpha$, giving

\begin{equation}\label{kinetic_equation_com}
    \begin{aligned}
        &(\partial_t - (\partial_t \boldsymbol{u}) \cdot \boldsymbol{\nabla_w} + (\boldsymbol{w} + \boldsymbol{u})\cdot (\boldsymbol{\nabla} - (\boldsymbol{\nabla u})\cdot \boldsymbol{\nabla_w})
        + \boldsymbol{F}_\alpha \cdot \boldsymbol{\nabla_w}) (f^M_\alpha + \delta f_\alpha)
         = \sum_\beta \bigg( C^T_{\alpha\beta}\{\delta f_\alpha\} + C^F_{\alpha\beta}\{\delta f_\beta\} \bigg),
    \end{aligned}
\end{equation}
where the non-linear part of the collision operator has been neglected.

The Maxwellian term may be evaluated identically to Ferziger and Kaper \cite{Ferziger_book} evaluating $(\mathcal{D} f_i)^{(0)}$, where the time derivatives acting on bulk quantities are replaced with the leading-order hydrodynamic (Euler) equations. Similarly, the $-(\partial_t \boldsymbol{u} + \boldsymbol{u} \cdot \boldsymbol{\nabla u})\cdot \boldsymbol{\nabla}_{\boldsymbol{w}}\delta f_\alpha $ term is rewritten using the Euler momentum equation $\partial_t \boldsymbol{u} + \boldsymbol{u} \cdot \boldsymbol{\nabla u} = - \frac{1}{\rho}\boldsymbol{\nabla}p + \sum_\beta c_\beta \boldsymbol{F}_\beta$. This transforms (\ref{kinetic_equation_com}) to
\end{strip}\begin{strip} 
\begin{equation}\label{kinetic_eqn_euler_subbed}
    \begin{aligned}
        &\biggl(\boldsymbol{w}\cdot \bigg[ \frac{n}{n_\alpha}\boldsymbol{d}_\alpha + \bigg( \frac{m_\alpha w^2}{2T} - \frac{5}{2} \bigg) \boldsymbol{\nabla}\ln T \bigg] + \frac{m_\alpha}{T} \bigg(\boldsymbol{w w} - \frac{1}{3}w^2 \mathbb{I} \bigg)\colon \boldsymbol{\nabla u} \biggl) f^M_\alpha
        \\&+\biggl(\partial_t  + (\boldsymbol{w} + \boldsymbol{u})\cdot \boldsymbol{\nabla} + (  -\boldsymbol{w} \cdot \boldsymbol{\nabla u} + \frac{1}{\rho}\boldsymbol{\nabla}p + \boldsymbol{F}_\alpha - \sum_\beta c_\beta \boldsymbol{F}_\beta ) \cdot \boldsymbol{\nabla_w} \biggl) \delta f_\alpha = \sum_\beta \bigg( C^T_{\alpha\beta}\{\delta f_\alpha\} + C^F_{\alpha\beta}\{\delta f_\beta\} \bigg).
    \end{aligned}
\end{equation}

Writing \ref{kinetic_eqn_euler_subbed} in 1D2V coordinates and neglecting the hydrodynamic velocity assuming $u \ll v_{\text{th}}$ yields the kinetic equation for $\delta f_\alpha$ as

\begin{equation}
    \label{kinetic_equation_for_deviation}
    \begin{aligned}
        \partial_t \delta f_\alpha &= \sum_\beta \bigg( C^T_{\alpha\beta}\{\delta f_\alpha\} + C^F_{\alpha\beta}\{\delta f_\beta\} \bigg) - \biggl[ \mu w \partial_z  
        + \bigg( \frac{1}{\rho}\partial_z p + F_\alpha - \sum_\beta c_\beta F_\beta \bigg)  \bigg( \mu \partial_w  + \frac{1-\mu^2}{w}\partial_\mu  \bigg) \biggl] \delta f_\alpha 
        \\&-\mu w f^M_\alpha  \bigg(\frac{n}{n_\alpha} d_\alpha + \bigg( \frac{m_\alpha }{2T}w^2 -\frac{5}{2}\bigg)\partial_z \ln T  \bigg)
        \,.
    \end{aligned}
\end{equation}

\section{Field-particle operator}\label{appendix_field_particle_operator}

The Legendre components of the field-particle operator \cite{10.1063/1.4936799} are

\begin{equation}\begin{aligned}
    C^{F,(l)}_{\alpha\beta}\{ \delta f^{(l)}_\beta \} &= \frac{2}{2l+1} \int_{-1}^1 d\mu\ P_l(\mu) C^{F}_{\alpha\beta}\{ \delta f_\beta \}
    \\&=   4\pi \bigg(\frac{q_\alpha q_\beta}{m_\alpha} \bigg)^2 \ln \Lambda_{\alpha\beta}\frac{m_\alpha}{T_\alpha}  f^M_\alpha \Biggl[ \frac{4\pi T_\alpha}{m_\beta} \delta f_\beta^{(l)} - \mathcal{H}^{(l)}_\beta  + \bigg(\frac{m_\alpha}{m_\beta} - 1 \bigg) w \partial_w \mathcal{H}^{(l)}_\beta + \frac{m_\alpha w^2}{2 T_\alpha} \partial_w^2 \mathcal{G}^{(l)}_\beta \Biggl],
\end{aligned}\end{equation} 
where

\begin{equation} \begin{aligned}
    &\mathcal{H}^{(l)}_\beta = \frac{4\pi}{2l+1} \biggl( \frac{1}{v^{l+1}}\int_0^w du\ u^{l+2} \delta f_\beta^{(l)}(u) + w^l \int_w^\infty du\ \frac{1}{u^{l-1}}\delta f_\beta^{(l)}(u)  \biggl)
     =  2\pi \biggl( \Psi_{\beta,1}^{(l)}(w) +  \Psi_{\beta,2}^{(l)}(w)  \biggl)
\end{aligned}\end{equation} 

\begin{equation} \begin{aligned}
    &w \partial_w \mathcal{H}^{(l)}_\beta = \frac{4\pi}{2l+1} \biggl( -\frac{l+1}{w^{l+2}}\int_0^w du\ u^{l+2} \delta f_\beta^{(l)}(u) + l w^{l-1} \int_w^\infty du\ \frac{1}{u^{l-1}}\delta f_\beta^{(l)}(u)  \biggl)
    =2\pi\biggl( l \Psi_{\beta,1}^{(l)}(w) - (l+1)\Psi_{\beta,2}^{(l)}(w)   \biggl)
\end{aligned}\end{equation} 

\begin{equation} \begin{aligned}
     w^2 \partial_w^2 \mathcal{G}^{(l)}_\beta 
     &= \frac{4\pi}{2l+1} \biggl( \frac{(l+1)(l+2)}{2l+3} \frac{1}{w^{l+3}}\int_0^w du\ u^{l+4} \delta f_\beta^{(l)}(u)- \frac{l(l-1)}{2l-1} \frac{1}{w^{l+1}} \int_0^w du\ u^{l+2} \delta f_\beta^{(l)}(u) 
     \\& + \frac{(l+1)(l+2)}{2l+3} w^l \int_w^\infty du\ \frac{1}{u^{l-1}}\delta f_\beta^{(l)}(u)- \frac{l(l-1)}{2l-1} w^{l-2} \int_w^\infty du\ \frac{1}{u^{l-3}}\delta f_\beta^{(l)}(u) \biggl)
    \\& = 2\pi \biggl( \frac{(l+1)(l+2)}{2l+3} (\Psi_{\beta,4}^{(l)}(w) + w^2 \Psi_{\beta,1}^{(l)}(w) )   - \frac{l(l-1)}{2l-1}  (\Psi_{\beta,3}^{(l)}(w) + w^2 \Psi_{\beta,2}^{(l)}(w) )  \biggl)
\end{aligned}\end{equation} 

where we have introduced the integrals

\begin{equation} \begin{aligned}\label{Psi_integrals}
    & \Psi_{\beta,1}^{(l)}(w) = \frac{2}{2l+1} w^l \int_w^\infty du\ u^{1-l} \delta f_\beta^{(l)}(u), \quad \Psi_{\beta,2}^{(l)}(w) = \frac{2}{2l+1} \frac{1}{w^{l+1}} \int_0^w du\ u^{2+l} \delta f_\beta^{(l)}(u),
    \\& \Psi_{\beta,3}^{(l)}(w) = \frac{2}{2l+1} w^l \int_w^\infty du\ u^{3-l} \delta f_\beta^{(l)}(u),\quad  \Psi_{\beta,4}^{(l)}(w) = \frac{2}{2l+1} \frac{1}{w^{l+1}} \int_0^w du\ u^{4+l} \delta f_\beta^{(l)}(u).
\end{aligned}\end{equation} 

Substituting into the field-particle operator, we have

\begin{equation} \begin{aligned}
    &C^{F,(l)}_{\alpha\beta}\{ \delta f^{(l)}_\beta \}  =  16\pi^2 \bigg(\frac{q_\alpha q_\beta}{m_\alpha} \bigg)^2 \ln \Lambda_{\alpha\beta} \frac{1}{v_{\text{th},\alpha}^{2}}  f^M_\alpha  \Biggl[ \frac{m_\alpha}{m_\beta} v_{\text{th},\alpha}^2 \delta f_\beta^{(l)}  + \bigg( - 1 + \bigg(\frac{m_\alpha}{m_\beta} - 1 \bigg) l + \frac{(l+1)(l+2)}{2l+3} \frac{ w^2}{v_{\text{th},\alpha}^2} \bigg) \Psi_{\beta,1}^{(l)}(w)
    \\& + \bigg( - 1 - \bigg(\frac{m_\alpha}{m_\beta} - 1 \bigg) (l+1) - \frac{l(l-1)}{2l-1} \frac{ w^2}{v_{\text{th},\alpha}^2}   \bigg) \Psi_{\beta,2}^{(l)}(w) - \frac{l(l-1)}{2l-1} \frac{1}{v_{\text{th},\alpha}^2} \Psi_{\beta,3}^{(l)}(w) +  \frac{(l+1)(l+2)}{2l+3} \frac{1}{v_{\text{th},\alpha}^2} \Psi_{\beta,4}^{(l)}(w)   
    \Biggl].
\end{aligned}\end{equation} 

\end{strip}

\section*{Data availability}

The data that support the findings of this study are available from the corresponding author upon reasonable request.

\newcommand{\newblock}{}

\bibliographystyle{iopart-num}

\bibliography{bib.bib}

\end{document}